\documentclass[
floatfix,
superscriptaddress,
nofootinbib,
amsmath,amssymb,
aps,
pre,
longbibliography,
notitlepage,
showkeys
]{revtex4-2}

\usepackage[left=2.5cm,top=3cm,right=2.5cm,bottom=3cm,bindingoffset=0.5cm]{geometry}
\usepackage{graphicx}
\usepackage{hyperref}
\usepackage{lipsum}
\usepackage{comment}
\usepackage{braket}
\usepackage{rotating}
\usepackage{booktabs}
\usepackage{physics}
\usepackage{centernot}
\usepackage[dvipsnames]{xcolor}
\usepackage{inputenc}[utf-8]
\usepackage{tikz-network}
\usetikzlibrary{matrix}
\usepackage{natbib}[squares, numbers]
\usepackage{tikz}
\usepackage{subcaption}

\usepackage{array}
\newcolumntype{C}[1]{>{\centering\arraybackslash}p{#1}}

\begin{document}

\title{Reconstructing supply networks\footnote{Corresponding author: francois.lafond@inet.ox.ac.uk. We would like to thank the participants of the Alan Turing Institute workshop on ``firm-level supply networks'' held in Cambridge in July 2023. L.M. and F.L. acknowledge funding from Baillie Gifford and the Institute for New Economic Thinking at the Oxford Martin School. D.G. acknowledges funding from the European Union - NextGenerationEU - National Recovery and Resilience Plan (Piano Nazionale di Ripresa e Resilienza, PNRR), project `SoBigData.it - Strengthening the Italian RI for Social Mining and Big Data Analytics' (https://pnrr.sobigdata.it), Grant IR0000013 (n. 3264, 28/12/2021).}}
\author{Luca Mungo}
\affiliation{Institute for New Economic Thinking at the Oxford Martin School, University of Oxford, UK}
\affiliation{Mathematical Institute, University of Oxford, UK}
\author{Alexandra Brintrup}
\affiliation{Institute for Manufacturing, Department of Engineering, University of Cambridge, UK}
\author{Diego Garlaschelli}
\affiliation{IMT School of Advanced Studies, Lucca, Italy}
\affiliation{Lorentz Institute for Theoretical Physics, University of Leiden, The Netherlands}
\author{Fran\c{c}ois Lafond}
\affiliation{Institute for New Economic Thinking at the Oxford Martin School, University of Oxford, UK}
\affiliation{Smith School of Enterprise and the Environment, University of Oxford, UK}

\maketitle

\section*{Abstract}
\noindent 
Network reconstruction is a well-developed sub-field of network science, but it has only recently been applied to production networks, where nodes are firms and edges represent customer-supplier relationships. We review the literature that has flourished to infer the topology of these networks by partial, aggregate, or indirect observation of the data. We discuss why this is an important endeavour, what needs to be reconstructed, what makes it different from other network reconstruction problems, and how different researchers have approached the problem. We conclude with a research agenda.
\\
\noindent  \textbf{Keywords}: link prediction; supply networks\\



\section{Introduction}
\label{sec:introduction}

Following the 2008 financial crisis, financial networks have been extensively studied by the complex systems community. For example, studying liabilities in banking networks has been key to developing the notion of systemic risk \cite{haldane_systemic_2011, bardoscia_physics_2021}, and explaining how certain types of interconnections may amplify the impact of isolated shocks. A key component of this research was the development of methods to reconstruct the network of interdependencies between financial institutions, which are not easily observable.

More recently, systemic failures on of the \textit{supply network} have captured the attention of the complex systems community, as researchers observed the impact of several significant failures, such as disruptions following the Great East Japan Earthquake in 2011, protective equipment shortages during the COVID-19 pandemic, supply shocks after the Suez Canal obstruction by the Ever Given, and the energy supply chain reorganization due to the war in Ukraine. 

Production networks, also known as ``supply chains'' or  ``supply networks'', consist of millions of firms producing and exchanging goods and services. From a mathematical perspective, they can be represented as weighted, directed graphs, where nodes symbolize firms (or establishments), and links may denote a supply-buy relationship with weights denoting transaction volume, such as the monetary value of the goods or services supplied over a given period.

Supply networks share many properties with other economic networks, but also exhibit unique features. Some of their empirical properties include \cite{bacilieri2023}:  small-world properties (short average path lengths and high clustering), heavy-tailed degree distributions, heavy-tailed (link and/or node) weight distributions, strong correlations between node strength and degree, and similarly between in- and out-degrees. It is also relatively well documented that, like biological and technological networks but unlike social networks derived from co-affiliation~\cite{newman2022}, supply networks feature negative degree assortativity.

However, supply networks are in many ways very different from other natural and economic networks. Their properties are deeply influenced by their function. First, the likelihood of a link between any two firms is driven by what the two firms are producing: for instance, steel manufacturers buy more iron than sugar. In general, each link in a supply network may represent one or more types of products; the diversity of products involved may depend on how the data are collected and may crucially affect network properties such as the reciprocity of connections. Product quality also plays a role, with ``high quality'' firms usually connecting with other ``high quality'' firms \cite{demir2023}. Second, supply networks are strongly embedded in geographic space, so that the likelihood of connections and their intensity decreases with distance \cite{bernard2019b}. Third, in contrast to financial networks, supply networks are less constrained by strict external regulations, and emerge as the result of a decentralized multi-criteria optimization process whereby millions of organizations simultaneously attempt to outsource in a way that minimizes their costs while maintaining acceptable levels of resilience to disruptions, for instance by multi-sourcing.

These characteristics make production networks incredibly complex: in modern economies, a sophisticated product such as an aircraft might involve contracting thousands of firms and sourcing millions of parts that cross national borders multiple times. Organizations in the network choose their dyadic relations and make local decisions, but hardly have visibility over their wider network. No single entity controls, designs and keeps track of the large-scale emergent network. Visibility over the network is, however, increasingly important for several reasons:  monitoring of environmental pledges to ensure firms quantify their greenhouse gas emissions, including those from their suppliers and customers; food and pharmaceutical traceability; analysing and improving supply chain resilience; and supply chain due diligence to ensure that actors that violate human rights or engage in environmentally damaging actions are not present in the chain.

In the past decade, researchers in economics and complex systems have worked extensively to better understand supply chains.  A key barrier to these studies has been a lack of data, as supply chains compete with one another \cite{christopher2011supply}, making information on them highly commercially sensitive. As a result, most studies to date have used firm-centred (e.g. starting with \cite{choi2002unveiling}) or sector-specific ({e.g. global automotive \cite{brintrup2016topological} and aerospace \cite{brintrup2015supply}, computer and electronics \cite{perera2017network}) supply chains}. 
While firm-centric and industry-specific studies have been important to gather insights into how network features shape the operation of supply chains, it remains hard to generalize these findings, due to the sector-specific and incomplete nature of these datasets.

Due to the above challenges, several recent studies have suggested the development of methods to reconstruct or predict the existence of hidden links in supply chain networks, offering a variety of approaches. These range from the use of natural language processing to extract and infer data from the World Wide Web to probabilistic maximum-entropy methods, each with varying success rates.

In this paper, we synthesize recent research on reconstructing supply networks. We start by describing the key problems: what data is available, what data is missing, and how to evaluate reconstruction performance (Section \ref{sec:what-is-missing}). We then summarise recent approaches to inferring the network topology (Section \ref{sec:topology}), and to infer the values of transactions when the topology is known (Section \ref{sub:links-weights}). We conclude with a discussion (Section \ref{sec:discussion}) and two research agendas (Section \ref{sec:agendas}) focusing on macroeconomic and supply chain management applications.

\section{The supply network reconstruction problem}
\label{sec:what-is-missing}

Production networks can be modelled at different levels of detail, both for nodes and edges. Naturally, the properties of the network depend on the level of aggregation. 

At the most granular level, nodes would represent individual production plants where goods undergo processing and transformation. A more aggregate model would equate nodes with the companies operating these plants. One could further aggregate by either consolidating firms under a common parent company or grouping them by industry sector\footnote{
One could think that the industry level is more aggregated than the firm. While this is mostly true, it is sometimes important to recognize that large firms span many industries. Indeed, industry-level input-output networks produced by National Accounts arise from Supply and Use Tables, which attempt to reallocate the output and inputs of multi-product firms into their appropriate categories.
}.

Firms exchange various goods and services. In a very detailed approach, each product type could be identified with a specific type of edge, rendering the production network as an edge-labelled multigraph. A simpler model would connect two nodes if they are involved in any type of trade, irrespective of the products' nature. Link weights can also have different definitions, measuring either the flow of goods (in terms, e.g., of the number of items traded) or the monetary value of such flow.

In the context of this paper, we define a \textit{supply network} $G$ as a graph where nodes represent firms while directed, weighted links represent the value of the flow of goods and services in a buyer-customer relation. This definition proves practical when reconstructing real-world supply networks from empirical data, which frequently adopts this format.

\subsection{What data is available?}

Almost all countries officially release Input-Output (I-O) tables, which provide the flow of money between industries, typically at the level of 50-500 industries. While we focus on firms here, this data is sometimes useful in the methods below. Besides, I-O tables provide a meso-scale ground truth that could be a good target for reconstruction methods.


\citeauthor{bacilieri2023} provides a taxonomy of existing datasets documenting different representations of supply networks. These are mainly: commercial datasets, confidential datasets held by governments, payment data, and industry-specific datasets. We briefly describe these types of data below.

Purchasing data from data providers, such as FactSet, Capital IQ, or Bloomberg is relatively straightforward, but commercial datasets can be very expensive, and cover only a fraction of firms, a very small fraction of links, and do not systematically include the value of the transactions. As commercial data providers typically assemble their data from publicly available information, researchers may also decide to collect this information themselves. An example is the extraction of data from the World Wide Web, after which machine learning algorithms are trained to predict supply-buy relationships \cite{wichmann2020}. Such an approach enables researchers to successfully gather rudimentary maps of supply chains, although it is limited to publicly available data, hence necessitating reconstruction efforts to identify missing relationships. 

The option of using government-held data necessitates datasets to be shared by national authorities, which may not always be feasible. However,  where data has been collected by a national authority it tends to be of very high quality. For example, VAT reporting may contain the value of transactions and timestamped data between virtually all firms within a country. \citeauthor{bacilieri2023} show that VAT datasets with no reporting thresholds exhibit strikingly similar properties, while incomplete datasets (either because of a reporting threshold or because they are assembled from publicly available information) usually have fewer links, so that many key statistics are likely to be highly biased.

A third option is payment data, which is usually (but not always) limited to individual banks collecting payment flows between their client firms (see, e.g., \cite{Ialongo2022}). Although it is not guaranteed that every transaction corresponds to a business link within a supply network, it can be viewed as a plausible indicator. These datasets are extremely detailed for any subset of firms affiliated with the same bank. However, they do not cover firms served by different banks or accounts held by their clients in different institutions.  

Finally, datasets focusing on specific industry verticals are also sometimes gathered by private companies (e.g., MarkLines' automotive dataset used in \citet{Brintrup2018})) and public regulatory bodies (e.g., the U.S. Drug Enforcement Administration's dataset of controlled substances flow). However, they are usually limited to specific geographies and production sectors.

There are no large-scale publicly available datasets on firm-level production networks, making it impossible at the moment to portray the global supply network. Summing up the number of nodes in the datasets reported in \citet{bacilieri2023} gives less than 3m, so less than $1\%$ of the 300m nodes reported earlier. Merging all the available datasets would give only an even smaller portion of the links and weights. This limitation forces researchers to use alternative options to proxy supply networks from smaller-scale, more specific datasets. These methodologies, developed to reconstruct or infer missing information about supply networks, are the main focus of this paper.

\subsection{A taxonomy of supply network reconstruction approaches}
\label{subsec:analogies-and-differences}

Clearly, what we actually mean by `reconstructing' a supply network necessarily depends on the data already available to the researchers and on the ultimate use of the (inferred) network, i.e. the goal of the analysis. We discuss these points in what follows and classify the studies we review along four primary axes. We do not see these classifications as having rigid boundaries, but rather as providing continuous dimensions along which models can be placed.

\paragraph{Predicting network topology and/or weights on transactions.} Consider a matrix $\Omega$ where $\Omega_{ij}$ shows the amount paid by $j$ to $i$. We distinguish between methods that focus only on finding the network's \textit{topology}, i.e., the presence or absence of a commercial connection between two firms encoded in the (binary) adjacency matrix $A_{ij}=1 \leftrightarrow \Omega_{ij}>0$, and those that assume that the adjacency matrix is known and try to infer the monetary value of the existing connections, i.e. the \textit{link weights} $\Omega_{ij}|A_{ij}=1$ (see also point \emph{c} below). Note that some methods try to simultaneously reconstruct both the topology and the weights of the network. Most of the methods we review focus on network topology. 

\paragraph{Predicting individual links or the full network.} Some methods focus on identifying the presence of specific links independently, while others try to reconstruct the entire network at once. The difference is subtle, yet important. Typically, links in real-world production networks are not independent. This happens, for instance, if firms tend to avoid ``multi-sourcing'': this happens if, when they are connected to supplier $j$ for a key input, they are less likely to be connected to other suppliers for that input.
In reconstruction methods, links are sometimes assumed to be mutually dependent, and sometimes assumed to be independent.
Generally (but not necessarily), the assumption made is related to the ultimate goal of the reconstruction method.
The task of trying to identify the presence of specific links is usually known as \textit{link prediction} \citep{lu2011}, while that of inferring the full network architecture is referred to (at least in this paper) as \textit{network inference}. In general, network inference computes the full distribution $P\left(G\right)$ over the set $\mathcal{G}=\{G\}$ of all possible networks. Link prediction, instead, computes the marginal probability $p_{ij}$ of an edge between nodes $i$ and $j$\footnote{
    More generally, link prediction methods produce a \textit{score} $s_{ij}$, such that $s_{ij} > s_{kl} \implies p_{ij} > p_{kl}$. However, such scores are not necessarily smaller than one, and the ratio between two scores is not necessarily equal to the ratio between links probabilities.
}. 
Again, there is no hard boundary between the two methods, which are occasionally equivalent: if one considers link independence as (the result of) a modelling assumption, computing the values $\{p_{ij}\}$ for all pairs of nodes and reconstructing the whole network become two  equivalent operations, as the probability $P\left(G\right)$ factorizes as
\begin{equation}
\label{eq:factorization}
    P(G) = \prod_{\left(i, j\right) \in E(G)} p_{ij} \prod_{\left(i, j\right) \notin E(G)} \left(1 - p_{ij}\right),
\end{equation}
where $E(G)$ denotes the set of edges realized in graph $G$. 
In this case, link prediction and network inference coincide.
On the other hand, whenever the full probability $P\left(G\right)$ in a network inference method is available (and irrespective of whether edges are assumed to be independent or not), it is always possible to compute the \emph{marginal} connection probability $p_{ij}$ as $p_{ij} = P\left(A_{ij} = 1\right) = \sum_{G \in \mathcal{G}} P\left(G\right)A_{ij}$ and use it in a link prediction exercise.

It is fair to say that the factorization in Eq.~\eqref{eq:factorization} is, at most, only approximately true in reality. However, some methods with independent edges can still capture meso- and macro-scale features of the network (see, e.g., \cite{Ialongo2022}) and, by framing the reconstruction problem as a binary classification task, link prediction facilitates easy comparison of methods through standard performance metrics.

\paragraph{Using topological information or not.} Of course, all reconstruction methods need, at the end of the procedure, the whole empirical network as the `ground truth' to \textit{test} their predictions. However, while some methods need the full adjacency matrix also in their training, other methods can learn from node-level or pair-level features only. This is important because the methods that do not rely on the adjacency matrix for training can be used in contexts where the detailed network is not observed, as long as certain node-level (and possibly pair-level) features are available.

\paragraph{Probabilistic or deterministic.} Some models produce \textit{deterministic} outputs, usually finding a network configuration by minimizing or maximizing a given loss function. Consequently, their output is a single network realisation that is on one hand optimal according to some score, but on the other hand very unlikely to represent the true network. 
Other methods provide \textit{probabilities} over possible network realisations. The goal of these methods can then be viewed as finding a `good' probability distribution, peaked `around' or `close' to the true one. Equipped with this probability distribution, researchers can find the typical and most likely realisations of the network and compute, for instance, expected values and confidence intervals for properties of the network.

\subsection{Evaluating the reconstructed networks}
\label{subsec:performance-metrics}

In their review paper on network reconstruction, \citeauthor{SquartiniEtAl2018} provide a useful taxonomy of performance metrics: \textit{statistical}, \textit{topological}, and \textit{dynamical} indicators.

\

\textit{Statistical} indicators evaluate the quality of the reconstructed network on a link-by-link (or weight-by-weight) basis. Different statistical indicators apply to deterministic and probabilistic outcomes.

In the realm of deterministic outcomes, perhaps the most commonly employed indicator is \textit{accuracy} (or \textit{precision}), the proportion of correct predictions. In supply networks, however, there is a strong class imbalance: the number of pairs not linked is much higher than the number of pairs linked. Thus, it is generally easy to make ``correct'' predictions since predicting that a link does not exist is very likely to be correct. For this reason, a commonly used metric is the \textit{F1-score}, defined as the harmonic mean of precision and recall (how many existing links are predicted as existing), which offers a more balanced performance metric in unbalanced datasets.

For probabilistic reconstructions, the evaluation is often based on the \textit{area under the receiver operating characteristic curve} (AUROC) and the \textit{area under the precision-recall curve}. AUROC, derived from the Receiver Operating Characteristic (ROC) curve, essentially quantifies the ablity of the models to discern between classes at varying threshold levels. The ROC curve plots the true positive rate (recall) against the false positive rate for different decision thresholds (i.e., by considering ``true'' all the predictions with probability larger than a certain threshold $\tau$, for different values of $\tau$), giving insights into the trade-off between sensitivity (true positive rate) and specificity (true negative rate). The AUROC, being the area under this curve, ranges from 0.5 to 1, with 1 implying an ideal classifier and 0.5 corresponding to no better than random guessing. 

Because statistical indicators focus on individual links, they may not adequately evaluate if the reconstructed network replicates complex network structures. \textit{Topological} indicators measure how well the network's macro-level and meso-level features are reproduced. Topological indicators gauge how effectively the reconstruction captures the network `coarse-grained' features. For instance, \citeauthor{Ialongo2022}, validate their reconstruction methodology by assessing how accurately it replicates the network degree distribution.

Topological indicators can tell us whether the reconstructed and true networks are ``similar''. However, ultimately the key question is whether a reconstructed network is good enough to give good answers to substantive economic questions. \textit{Dynamical} (or more generally model-based) indicators assess the similarity in the process' evolution on the real and reconstructed networks. As an example, \citeauthor{diem2022quantifying} introduced the \textit{Economic Systemic Risk Index} (ESRI) to quantify each firm's importance within an economy. The metric measures the percentage drop in the economy's overall production caused by the removal of a firm from the network. Its computation requires running a dynamical process, wherein the sudden disappearance of a firm first impacts its suppliers and customers and, iteratively, spreads to firms that are further away in the network, until the system reaches an equilibrium. 
Conceivably, accurately estimating firm-level ESRI may only necessitate identifying a subset of key links, so a good prediction of the other links is not necessarily important for the final economic result.

Armed with these evaluation indicators, we now examine in detail the models employed for reconstructing production networks, starting from methods focusing only on the network topology, and then discussing methods for reconstructing network weights.

\section{Reconstructing the network topology}
\label{sec:topology}

We start by reviewing studies that reconstruct the network using link prediction, and then those that do so using network inference methods. Table \ref{tab:topology-reconstruction-resume} provides an overall summary of the methods and their differences.

\begin{table}[ht]
\centering
\footnotesize
\renewcommand{\arraystretch}{1.4}
\begin{tabular}{r>{\raggedright\arraybackslash}p{1.7cm}>{\raggedright\arraybackslash}p{25mm}>{\raggedright\arraybackslash}p{4cm}c}
    \toprule
        & \textit{Coverage} & \textit{Dataset} & \textit{Inputs} & \footnotesize{\textit{Probabilistic}} \\
    \midrule
    \citet{mori2012} & Regional & Tokyo Area Manufacturing Firms, Source unspecified & Several features regarding firms' activities, balance sheets, management & \\
    \citet{zuo_kajikawa_mori2016} & National & Tokyo Shoko Research & Firms' sales, profits, industrial sector, location, number of employees, network centrality & \\
    \citet{sasaki_sakata_2017} & Regional & Tohoku region, Teikoku Databank & Firms' sales, capital, size, industrial sector, network centrality & X \\ 
    \citet{lee2022} & National & Korean Enterprise Data & Description of firms' activities, firms' industrial sector and location, aggregate transaction volumes between industrial sectors & X\\
    \citet{Brintrup2018} & Automotive & Markline Automotive Information Platform & Firms' known connections, products, intermediate inputs & X \\
    \citet{Kosasih2022} & Automotive & Markline Automotive Information Platform & Firms' known connections  & X \\
    \citet{minakawa_et_al_2023} & Global & Asian bank's transaction data & Firms' known connection, description of firms' activities & X \\
    \\
    \citet{mungo2023} & Global, National & Compustat, FactSet, Ecuador VAT & Firms' sales, industrial sector, location & X \\
    \citet{zhang_et_al_2012} & Global & Specialized Press (Reuters) & Media coverage & X \\
    \citet{wichmann2020} & Global & Specialized Press & Media coverage &  \\
    \citet{bert2023}& Global & Specialized Press & Media coverage  &  \\
    \citet{reisch2022inferring} & National & Phone calls, survey, Hungary VAT & Firms' phone calls, national IOTs  & X \\
    \citet{Hooijmaaijers2019} & National, 4 commodity groups. & IOTs, Business Register, Structural Business Statistics  &  Firms' known connections, sales, geographic location, industrial sector  & \\
    \citet{hillman2021} & National & IOTs, Business Register, Structural Business Statistics  &  Firms' known connections, sales, geographic location, industrial sector  & \\
    \citet{Ialongo2022} & National & Dutch banks' transaction data  & Firms' sales, intermediate expenses by sector, network density (for calibration)  & X \\
    \citet{mungo2023revealing} & Global & FactSet  &  Firms' sales (time series), industrial sector, network sectoral structure & \\
    \bottomrule
\end{tabular}
\caption{Overview of the papers that reconstruct the supply network topology.}
\label{tab:topology-reconstruction-resume}
\end{table}

\subsection{Link prediction}
\label{subsub:link-prediction}

\subsubsection{Setting up the problem}

\begin{figure}
    \centering
    \begin{subfigure}{0.45\textwidth}
        \includegraphics[width=\textwidth]{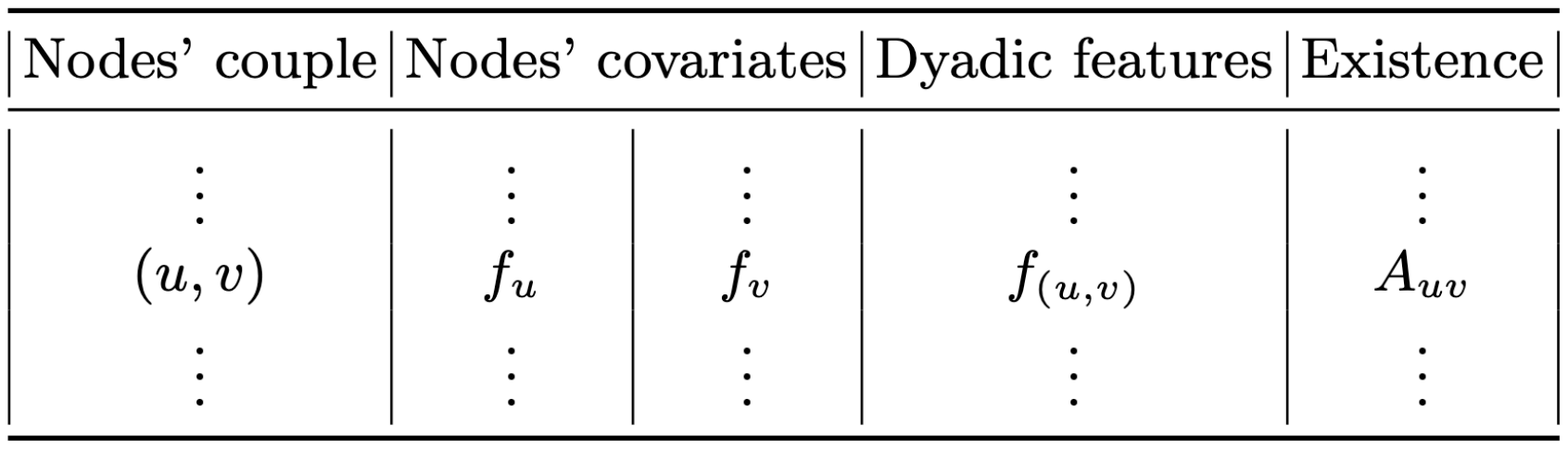}
        \caption{}
    \end{subfigure}
    \hfill
    \begin{subfigure}{0.45\textwidth}
\begin{tikzpicture}

\def\rectOneHeight{4}
\def\rectTwoHeight{2}

\definecolor{myBlue}{RGB}{0,0,255}
\definecolor{myRed}{RGB}{255,0,0}

\fill[myRed] (-3, -3*\rectOneHeight/4) rectangle (-2,0);
\fill[myBlue] (-3,0) rectangle (-2, \rectOneHeight/4);

\fill[myRed] (2,-\rectTwoHeight) rectangle (3,-1);
\fill[myBlue] (2,-1) rectangle (3, \rectTwoHeight/2-1);

\draw[-{Latex[width=3mm,length=3mm]},gray,line width=1pt] (-1.7, -1) -- node[above] {Undersampling} (1.7, -1);

\end{tikzpicture}
   
        \caption{}
    \end{subfigure}
    \caption{(a) Datasets for link prediction are usually built by filling rows with two nodes features ($f_u$, $f_v$, $f_{u, v}$) and by indicating if there is a link between the two nodes ($A_{u,v}$). (b) These datasets are usually undersampled: in the original dataset, a small minority of the rows will be s.t. $A_{u, v} = 1$ (blue), while most of the rows will be s.t. $A_{u, v} = 0$ (red); undersampling discards a portion of them to generate a more balanced dataset.}
    \label{fig:dataset}
\end{figure}

An early stream of research employs machine learning for link prediction in production networks. The key idea is to construct a dataset in the form of Fig.~\ref{fig:dataset}A, where for each pair $\left(i, j\right)$, we collect some features $f_{(i,j)}$ that can be features of each node (e.g., the product it makes, its total sales, etc.) or of the pair (e.g. geographical distance, whether they have a common supplier or client, etc.), and the response $A_{ij}$, which is equal to 0 or 1.

With such a dataset, one can then train a machine-learning classifier on a set of examples $\left\{f_{(i,j)}, A_{ij}\right\}$. Different papers have then made different choices for the predictors $f_{(i, j)}$ and the predictive algorithm, as we will discuss in detail. But before, let us note
another critical element, which is the construction of the dataset. Production networks are very sparse \cite{bacilieri2023}, so the ratio between the number of existing ($A_{ij} = 1$) and non-existing ($A_{ij} = 0$) links is very large. Therefore, training a model on the entire set of available examples might simply be computationally intractable (there are $\sim n^ 2$ pairs). Moreover, sampling a random subset would usually lead to poor predictions, because the scarce number of positive examples hinders the model's ability to effectively discriminate between the two classes. This phenomenon, known as the \textit{class imbalance} problem, can potentially lead to models that are biased toward predicting the majority class, thus failing to accurately identify the existing links.

This problem is commonly addressed by applying \textit{undersampling} (Fig.~\ref{fig:dataset}B), a technique that aims to rebalance the class distribution. In the context of production networks, undersampling involves carefully curating the training set to ensure a predetermined ratio between positive ($A_{ij} = 1$) and negative ($A_{ij} = 0$) examples. This controlled selection helps foster a more balanced, discriminative model and was employed in all the machine learning approaches that we are now set to survey. 

However, this procedure has implications for model evaluation. Typically, an algorithm is trained on a subsample (the training set), and evaluated on the remaining data (the testing set). If subsampling is done before the split into a testing and training set, the testing set will contain many more positives than a ``real-life'' testing set, so metrics such as accuracy will be severely biased. \cite{mungo2023} found that metrics such as AUC were not substantially affected by the undersampling ratio, so we will tend to report AUCs, which are more comparable across studies. Many studies, however, report the F-score, which is highly dependent on class imbalance \cite{mungo2023}, so when reporting F-scores we will also report undersampling ratios.

\subsubsection{Predicting new business partners}
\label{subsec:predicting-new-partners}

Interestingly, link prediction in production networks has not been originally pursued to reconstruct existing networks, but rather to build recommender systems that could suggest new partnerships to companies trying to expand their supplier or customer bases. In this framework, the ability of a model to identify existing (or past) supply-chain links is a target in so far as it is a proxy for their ability to make sensible recommendations, i.e., to identify \textit{candidate} links that firms could turn into existing ones.

Despite aiming for different goals, these studies share several similarities with those on network reconstruction in the problem's layout, framed as a link prediction task, and the tools used, often relying on statistical models and network science.

\citeauthor{mori2012} focuses on $\sim$ 30k manufacturing firms in Japan. They build a business partner recommendation system by feeding a Support Vector Machine (SVM) with several companies' features, such as size, industrial sector, and geographic location. On a dataset comprising $\sim$ 34k links and an equal number of negative instances, they achieve an F-score of $0.85$. The approach is refined in \cite{zuo_kajikawa_mori2016}, who still use an SVM but add topological properties in the list of companies' features, such as their degree, betweenness centrality, and closeness centrality. For a network of 180k firms and half a million links assembled through the Tokyo Shoko Research dataset, and again an undersampling ratio of 1:1, they achieve an F-score of $0.81$. 

\citeauthor{sasaki_sakata_2017} explicitly incorporate the network of second-tier suppliers and their respective industries, providing a more contextual analysis. The authors' intuition is that two firms within the same industry but with different suppliers will have different probabilities to sell to a specific customer. In other words, establishing a relationship between firms $A$ (supplier) and $B$ (customer) does not depend solely on the identity of $A$ and $B$, but also on who are $A$'s suppliers. Thus, the authors first extract from their network all the triads of firms connected in sequence (i.e., all the motifs $A\rightarrow B\rightarrow C$). Then, they replace each firm with its industrial sector (e.g., if we call $S_i$ the industrial sector of firm $i$, the triplet $A\rightarrow B\rightarrow C$ becomes $S_A\rightarrow S_B \rightarrow S_C$), and use a Bayesian model called \textit{n-gram} to compute the link probability between $B$ and $C$ given $B$ and $C$'s industrial sectors and the industrial sectors of $B$'s suppliers. Finally, the authors use these probabilities as features in a random model classifier, together with a few firms' attributes (total revenues, number of employees, etc.) and network centralities. The authors focus on $\sim$ 50k links in a network of 130k Japanese firms\footnote{
The authors test their method on ``new'' links, missing from their 2010 snapshot of the network and present in the 2011 snapshot. The data is provided by Teikoku Databank Ltd., a business intelligence company.
}, achieving an F-Score of $0.80$ with an undersampling ratio of 1:1.

More recently, \citeauthor{lee2022} integrated information on firms' geographical position and industrial sector with aggregate trade volumes between sectors and textual information on companies' activities and products. The authors encode this information and use it to train a deep neural network. On a sample of $\sim$ 90k connections between South Korean firms, where $20\%$ of the examples are used as a test set, the authors achieve an AUROC of $0.92$\footnote{The authors do not specify the undersampling ratio of their exercise.}.

This trajectory of studies reflects a consistent evolution in methodology, with each iteration contributing incremental enhancements in feature integration and model sophistication, partially akin to what we will see now for papers which address supply network reconstruction specifically.

\subsubsection{Can a firm better understand its supply network dependencies?}

From a supply chain management perspective, a focal firm is interested in understanding hidden dependencies within its supply network \-- for instance, two suppliers may rely on a hidden ``second tier'' supplier, creating a vulnerability for the focal firm that is invisible at first sight. In such a context, the focal firm would typically see a fair part of the network and could use this topological information to make further inferences. 

This is the context of the early investigation by \citeauthor{Brintrup2018}\cite{Brintrup2018}, who focuses on the supply networks of three specific major car manufacturers (Jaguar, Saab, and Volvo, using data from the Marklines Automotive Information Platform). Using their domain expertise, the authors create four features for each potential link $\left(i, j\right)$: \textit{Outsourcing Association} (the overlap between the goods produced by company $i$ and those bought by company $j$), \textit{Buyer Association} (how frequently firms that purchase the same inputs as firm $i$ also buy the products of firm $j$), \textit{Competition Association} (the overlap between the products of firm $i$ and those of firm $j$.), and \textit{Degrees} (the number of partners of each firm). Training a logistic regression and a Naive Bayes using these features yields an AUROC of around $0.8$.

In a subsequent paper \cite{Kosasih2022}, the authors refine their approach using Graph Neural Networks (GNNs) \cite{hamilton2020}. The concept underlying GNNs is that the network's topological information should not be distilled by the researchers through the design of specific features (as was the case with the association measures of the previous paper), but should instead be discovered automatically by the neural network. For production networks, the intuition is that the association measures designed in earlier work \cite{Brintrup2018}, while informative, might not convey all the information lying in the network's topology. Instead, a neural network provided with a sufficient amount of examples would identify patterns hidden to the researchers.

Practically, this is accomplished by: 1) for each link $l = \left(i, j\right)$, isolating subnetworks $G_{i}$, $G_{j}$ composed by the nodes $i$ and $j$, along with the set of their neighbours; 2) embedding each node $u$ in the subnetwork $G_{l} = G_i \cup G_j$ in a vector $f_{u, l}$\footnote{
    The embedding usually consists of computing an average distance $d$ between node $k$ and the nodes $i$ and $j$, and then embedding $k$ in a vector $f^k_{ij} = \delta_{dd'}$. The dimension of this vector is the maximum possible distance, which must be specified as a parameter of the model.
}; 3) feeding the nodes' embeddings $f_{u, l}$ to a series of $K$ \textit{graph convolutional layers}, which are nonlinear functions $f^{k+1}_{ul} = \phi\left(f^{k}_{ul}, \left\{k_u\right\} \right)$, where $k_u$ are the degrees of the nodes in $G_u$; 4) averaging the final vectors $f^K_{u, l}$ across all the different nodes $u$, generating an embedding vector $f'_{l}$ for the subnetwork $G_{l}$; 5) feeding the embedding through a sequence of fully connected layers to generate a single prediction for the probability $p_{ij}$.

The weights in the graph-convolutional and fully connected layers are trained with the usual backpropagation algorithm. The authors find a significant improvement compared to the previous approach, with the GNNs scoring an AUROC value $\sim 0.95$. While this is an impressive improvement in performance, a downside of this approach is that it becomes very difficult to interpret the predictions made by the neural network and develop novel insights into how firms connect.

A similar approach is proposed in \cite{minakawa_et_al_2023}, where the authors train a graph neural network with topological information and textual information on firms' activities, encoded via the Doc2Vec algorithm \cite{doc2vec}. On a network of 170k firms and 1.2M edges provided by a large Asian bank, the authors AUROC of 0.94-0.95, depending on the respective sizes of the training and the test data. They do not report the undersampling ratio.

\subsubsection{Predicting the supply networks of entire countries where no network data exist}

\citeauthor{mungo2023} use similar methods for a different purpose. They observed that in some countries,  excellent data is available, while in other countries (including the US), there is no fully reliable information on firm-to-firm transactions, creating a need for methods that predict the supply network using only information available locally (\citet{Hooijmaaijers2019}, reviewed in Section \ref{subsec:max-ent-weights}, first developed a method based on data held by the most statistical offices). Based on this observation, they ask whether a model trained on the production network of a country $A$ accurately predicts links within firms in another country $B$. 

In all countries, there is usually good data available on key features of firms and pairs of firms that could determine link formation. For example, it is well established that large firms have more connections \cite{bacilieri2023}, prefer to trade with geographically closer firms \cite{bernard2019b, bernard2019}, and have production recipes that put significant constraints on the inputs they buy. 
Based on these hypotheses, for each candidate link, the authors build a vector $f_{(i,j)}$ containing information on firms' sales, industrial sector, and geographical distance. They then train a \textit{gradient-boosting} model to predict link probability. 

The study is run on three different datasets: two commercial, global datasets (\textit{Compustat} and \textit{FactSet}) and one dataset covering (a subsample of) Ecuador's national production network, assembled by Ecuador's government using VAT data. When tested on the same dataset used to train the model, the approach scores an AUROC similar to that of the previous approach (from $\sim0.91$ to $\sim0.95$ depending on the dataset), suggesting that indeed, knowing a firm's products, location and size provides sufficient information to make decent predictions.

For making predictions on unobserved countries, they conduct two tests. In the first test, they considered different countries in the same dataset, for instance training their model on FactSet's US and Chinese networks and predicting links in Japan. In this case, the approach still performs relatively well (AUROC $> 0.75$). In the second test, they predict the links in Ecuador using FactSet and the other way around. Here, the performance deteriorates substantially, which the authors explain by showing that the distribution of features in FactSet (an incomplete, commercial dataset with large firms in rich countries) and Ecuador (a complete administrative dataset, with all firms from a developing economy) are very different. 

This partial success suggests that there is potential for further studies, but using multiple administrative datasets. For instance, while it is not possible to predict the Ecuadorian administrative data using the commercial data from FactSet, it might still be possible using similar administrative datasets, given the results from \cite{bacilieri2023} showing that administrative datasets exhibit strikingly similar topological properties. This is a straightforward approach to reconstructing the global firm-level production network, using training data from a few countries, and large-scale firm-level datasets such as ORBIS.

\subsubsection{Leveraging alternative data: news and phone calls}
\label{subsub:beyond-standard}

The idea in \cite{zhang_et_al_2012} and \cite{wichmann2020} is that significant commercial deals might be announced in press releases or covered by the specialized press. 

\cite{zhang_et_al_2012} build a system to automate the analysis of articles and investor comments coming from Reuters and identify collaborative\footnote{Note that, for the authors, a ``collaborative relationship'' has a broader meaning than supply relationship. } and competitive relationships between companies. The authors web-scrape a corpus of $\sim 125k$ documents and manually annotate a sample of $4.5k$, overall identifying 505 relationships. Then, they use a Latent Dirichlet Allocation (LDA) algorithm (a widely used algorithm in text analysis) to examine these examples, finding that the algorithm identifies collaborative relationships with an AUROC of $0.87$.

Similarly, \cite{wichmann2020} automates the analysis of textual data (coming from Reuters Corpora TRC2 and RCV1, NewsIR16, and specific web searches) to find mentions of commercial deals between the firms. First, the authors collect a text corpus describing the relationships between firms. Then, they classify these relationships as either a commercial relationship (e.g., firm $i$ supplies firm $j$), an ownership relationship (firm $i$ owns firm $j$), or none of the previous. The annotated examples are embedded into numerical vectors using the word embeddings in the Glove dataset and finally used to train a Natural Language Processing (NLP) classifier with a BiLSTM architecture. $30\%$ of the sentences were left out of the data and used to assess the performance of the model, which scores an F1-score of $0.72$ with a class imabalance of 1:7. Unfortunately, the choice of evaluating the score on a binary metric (the F1-Score) does not allow a straightforward comparison with the previous approaches. However, the authors report that a random classifier would get an F1-Score of $0.38$. In a follow-up paper \cite{bert2023}, the authors improve their results by running the same study using a BERT model, and reach an F1-Score of $0.81$.

In \cite{reisch2022inferring}, instead, the authors use phone calls between companies and survey data to track down supplier-customer relationships in an undisclosed European country. The survey asked companies to list their ten most important suppliers and customers. On this subsample of the network, the authors find that if the average daily communication time between two firms $i$ and $j$,  denoted $\tau_{ij}$, is greater than $30$ seconds, the probability that these two firms are connected is $p_{ij} \approx 0.9$. Equipped with this observation, the authors reconstruct the network by first assuming the presence of a link between $i$ and $j$ if $\tau_{ij} > 30s$ and then assigning a direction to the link stochastically with a probability 
$$
p\left(i\rightarrow j\right) = \frac{\omega_{a_i b_j}}{\omega_{a_i b_j} + \omega_{b_ja_i}},
$$
where $a_i$ and $b_j$ are $i$ and $j$'s respective industrial sector, and $\omega_{ab}$ is the total amount of trade (in monetary value) from firms in sector $a$ to firms in sector $b$, as reported in the country's Input-Output tables\footnote{
A consequence of the algorithm choosing edge direction is that the reconstructed network has null reciprocity, while we know that real networks exhibit reciprocity of around a few percent \cite{bacilieri2023}.
}. The authors do not provide any `standard' evaluation metric for their reconstruction. However, they mention that choosing a threshold $\tau_{ij} = 30s/d$ minimizes the Kullback-Leibler divergence between the degree distribution of the reconstructed network and the degree distribution of a well-studied network, the Hungarian production network. The authors' ultimate goal was to compute firms' Economic Systemic Risk Index (ESRI, see Section \ref{subsec:performance-metrics}) in the reconstructed network, and they do find a good qualitative agreement between the ESRI sequence of firms in the reconstructed and the Hungarian network.

\subsection{Network Inference}
\label{subsub:network-inference}

A second stream of research tries to reconstruct the production network as a whole rather than link-by-link. We distinguish three sets of approaches: matching algorithms, maximum entropy methods, and methods based on time series correlations.

\subsubsection{Matching algorithms}
\label{subsub:matching}

A couple of papers have used matching algorithms to create supply networks. We classify these under ``Network Inference'' because while they reconstruct the network link-by-link, they typically try to match aggregate constraints, taken from I-O tables and/or from meso-level statistics published independently.

An early study is the one from Hooijmaaijers and Buiten (\cite{Hooijmaaijers2019}, see \cite{rachkov2021} for details), who devise an algorithm that matches firms based on commonly observable firm characteristics (industry, size, location) and I-O tables.

Roughly speaking, their method works as follows. First, using a relationship between sales and degrees of $s_i \propto k_i^{1.3}$ \cite{watanabe2013relations}, they can estimate out-degrees based on total sales. For expenses, using the I-O tables they can estimate expenses of each firm by industry, and assuming that in-degree by industry is a (specific) increasing function of expenses by industry, they can estimate the number of industry-specific suppliers for each firm. 

Knowing the degrees of all firms, the next task is to match them. To do this, they create pairwise scores based on assumptions about what determines the likelihood of a match. The final score is a linear combination of three scores: one that increases with firm size, one that decreases with distance, and one that acts as a bonus or penalty if the firms are in industries that trade in I-O tables. The matching algorithm then starts with the buyer that has the highest purchasing volume and goes in descending order. The number of suppliers connected to each buyer is determined by the buyer's in-degree. Among the potential suppliers, those with the highest scores are considered the most likely to trade with the buyer. If any of these top-rated suppliers have no remaining outgoing links, then the next most likely supplier in line is considered instead.

\citeauthor{hillman2021} introduced another algorithm, driven by their need to create a synthetic firm-level network for their agent-based model of the impact of the Covid-19 pandemic. Again, their method makes use of I-O tables and data on sales, although it does not use location information. Their algorithm is less clearly documented, but essentially works by first using I-O tables to determine which industries a firm should sell to, then allocating chunks of its sales to randomly selected firms in the buying industry. They show that their algorithm is able to reproduce a positive strength-degree relationship.

\subsubsection{Maximum-entropy for network inference}
\label{subsub:maximum-entropy-binary}

In a sense, matching algorithms try to distribute connections ``randomly'', while matching some aggregate properties of the network. However, to do so they introduce ``plausible'' assumptions, such as specific functional forms to create scores. Instead of introducing assumptions, the Maximum Entropy assigns a probability to each possible network in a ``maximally non-committal'' way. This leads to the question of whether introducing assumptions about what is not fully known is better than just maximizing entropy conditional only on what is fully known. This is the question of \citeauthor{rachkov2021}, who showed that the networks obtained from the matching method proposed in Ref. \cite{Hooijmaaijers2019} have different properties than those obtained using a simple maximum-entropy model, suggesting possible biases in heuristics-based reconstructions. That being said, simple maximum entropy methods are not well-suited for complete supply networks (i.e., not commodity-specific), because they do not use information on firms' products, which we know is a critical determinant of their probability to link.

\citeauthor{Ialongo2022} introduced a method that tackles this issue and simultaneously reconstructs the whole network topology and link weights (see Sec.~\ref{sub:links-weights} for the weights). Following a long-standing tradition in network reconstruction \cite{SquartiniEtAl2018}, they compute a probability distribution $P(G)$ over the set of all possible graphs $\mathcal{G}$ that maximizes the Shannon Entropy $\mathcal{S}$,
$$
    \mathcal{S} = -\sum_{G \in \mathcal{G}}P\left(G\right)\ln P\left(G\right).
$$
The maximization is subject to a normalization constraint, $ \sum_{G \in \mathcal{G}} P(G) = 1$, and a collection of constraints $\boldsymbol{\tilde{c}}$ representing the macroscopic properties enforced on the system. These constraints are usually enforced in a soft way, that is, by constraining the expected values of the constraints over the set of all possible networks $\mathcal{G}$,
$$
\sum_{G \in \mathcal{G}} P(G)c_i\left(G\right) = \tilde{c}_i.
$$

The authors expand on a pre-existing model \cite{parisi2020}, constraining the network's density $\rho$, each firm's total sales $\omega^{out}_{i}$ and the money spent by firm $i$ on inputs from each industrial sector $a$, $\left\{\omega_{a\rightarrow i}\right\}$. 
However, as we have already emphasized, a crucial feature in supply networks is that firms connect to others specifically for the products they make. A method that does not take into account the product or industry of the firm is, in the context of supply networks, doomed to fail.

As a result, the authors design a new model able to handle sector-specific constraints. For instance, in a hypothetical economy with two sectors, $a$ and $b$, the model enforces three constraints on each firm: one for total sales, $\sum_{G \in \mathcal{G}} P\left(G\right)\omega^{out}_{i} = \tilde{\omega}^{out}_i$ and one for spending on each of the sectors:  the money spent on inputs from sector $a$, $\sum_{G \in \mathcal{G}} P\left(G\right)\omega_{a\rightarrow i} = \tilde{\omega}_{a \rightarrow i}$, and the spending on inputs from sector $b$, $\sum_{G \in \mathcal{G}} P\left(G\right)\omega_{b\rightarrow i} = \tilde{\omega}_{b \rightarrow i}$ (we use tildas to denote observed quantities). The model accepts an analytical solution for the marginals $p_{ij}$,
\begin{equation}
\label{eq:p_ij-stripe-max-ent}
    p_{ij} = \frac{z\tilde{\omega}^{out}_i\tilde{\omega}_{a_i \rightarrow j}}{1 + z\tilde{\omega}^{out}_i\tilde{\omega}_{a_i \rightarrow j}},
\end{equation}
where $a_i$ is the industrial sector of firm $i$, and $z$ is chosen such that $\sum_i\sum_{j \neq i} p_{ij} = \tilde{\rho}$. 

The authors show that their method significantly improves upon the model by \cite{parisi2020}, where each firm is subject to a single constraint for the overall intermediate expenses. In a maximum-entropy framework, imposing only one constraint on the intermediate expenses would distribute a firm's supplier equally across all industrial sectors. This is at odds with the reality of supply chains, where firms require only a select range of goods from the basket of products available in an economy. 

The authors do not report any standard reconstruction metric, but they show that the in-degree and out-degree distribution of the reconstructed network are, in expectation, in good agreement with the empirical degree distribution. Moreover, the relationship between degrees and strengths of firms is generally well replicated.

A limitation of all the studies discussed so far is that they consider only firm-to-firm links. For macroeconomic applications, it would be useful to reconstruct complete synthetic populations (see Sec.~\ref{sec:agendas}), including links between firms (including banks) and consumers/workers. \citeauthor{hazan2019maximum} uses a maximum-entropy approach (more precisely, the fitness-induced configuration model, \cite{garlaschelli_loffredo_2004}) for firm-to-firm networks and firm-to-consumer networks, taking average degrees from the literature to estimate $z$ separately in each network.

\subsubsection{Leveraging the correlation matrix using graph learning}

An established literature tackles the problem of reconstructing a network starting from $N$ node-level time series encoded in vectors $x\left(t\right) \in \mathbb{R}^N$ \cite{dong2019, peixoto2022}. The general philosophy is that the structure of the network $\mathcal{G}$ determines the joint probability distribution of the observations. If one assumes that each observation $x\left(t\right)$ is drawn from a probability distribution $p\left(x|\Theta\right)$ with a parameter matrix $\Theta \in \mathbb{R}^{N\times N}$, the problem of reconstructing a graph, or \textit{graph learning}, becomes that of finding the correct value of $\Theta$.

Production networks serve as a contagion channel for economic shocks. They spread negative or positive shocks from one firm to its customers and suppliers, generating correlations between firms' fundamentals, such as market valuation and sales \cite{barrot2016, Carvalho2019,carvalho2021}. Starting from this observation and leveraging the graph learning literature, \citeauthor{mungo2023revealing} introduce a method to reconstruct the production network from the time series of firm sales, $s_i\left(t\right)$. First, the authors show empirically that the correlation between the log-growth rates of firms connected in the production network surpasses the average correlation yielded by randomly sampled firm pairs, and this excess correlation decreases as firms get further apart in the supply chain. Then, the authors harness this observation to design a network reconstruction approach, framed within Gaussian Markov Random Fields \cite{dong2019}. Adapting a modern graph learning strategy \cite{Kumar_et_al_2019}, the authors assumed that the growth time series data could be modelled as a sequence of draws from a multivariate Gaussian distribution. This distribution's precision matrix (the inverse of the covariance matrix) is, in turn, identified with the network Laplacian $L = D - A$ where $D_{ij} = k_i\delta_{ij}$. To estimate the precision matrix, the authors employed a maximum likelihood approach, constraining the possible Laplacians $L$ to preserve the expected connections' density within and across economic sectors. In addition, a penalization term is included to enforce network sparsity.

Upon assessment against smaller network fragments, their methodology reports an F1-score within the range of $0.2-0.3$. Nevertheless, it does not consistently surpass all benchmark tests under consideration. While it is true that, on average, firms that are more closely connected are more correlated, there is a lot of overlap between the distributions of correlations at various distances. In other words, knowing that firms are highly correlated is not very informative of their distance, making the task of network inference based on time series data very challenging.

\section{Inferring the value of transactions}
\label{sub:links-weights}

While methods for reconstructing weights have been used extensively on financial and global trade networks \cite[e.g.][]{anand2018, SquartiniEtAl2018, cimini_mastrandrea_squartini_2021} and aggregate I-O tables \cite[e.g.][]{golan1994}, their application to firm-level networks is relatively novel. A first set of methods uses meso-level information from I-O tables, while another set of papers relies on the maximum entropy principle.

\subsection{Matching I-O tables}
\label{subsec:prop-weights}

\citeauthor{InoueTodo2019} incorporate aggregate I-O information into their estimates of the weights in the supply network of Japan. They assign to each link between a supplier $i$ and a customer $j$ a weight proportional to the product of firm sales, $\omega_{ij} \propto \tilde{\omega}^{\text{out}}_i\frac{\tilde{\omega}^{\text{out}}_j}{\sum_{j \in \mathcal{N}_i} \tilde{\omega}^{\text{out}}_j}$, where $\sum_{j \in \mathcal{N}_i}$ means that the sum runs only on $i$'s customers. The weights are then rescaled to align with the aggregate transaction amounts within industry sectors $\tilde{\omega}_{ab}$,
$$
    \omega_{ij} = \tilde{\omega}^{\text{out}}_i \frac{\tilde{\omega}^{\text{out}}_j}{\sum_{j \in \mathcal{N}_i} \tilde{\omega}^{\text{out}}_j}\frac{\tilde{\omega}_{a_i b_j}}{\sum_{k \in a_i, l \in b_j} \tilde{\omega}^{\text{out}}_k\tilde{\omega}^{\text{out}}_l},
$$
where $a_i$ and $b_j$ denote the respective industrial sectors of $i$ and $j$.
A similar approach has been used by \cite{hillman2021} where, starting from data on firms' sales and inputs, the authors construct individual-firm networks, that, when aggregated, align with the sectoral IO table. The authors rescale firms' input and output to match the IO tables\footnote{
More precisely, they match intermediate inputs (roughly, inputs that are neither labour nor investment goods), and gross output (roughly, total sales).
}, and then allocate links in the network with an iterative algorithm that matches buyers to suppliers, while also imposing that larger firms will have more customers. The weight of each connection is then set to the smallest value between the supplier's maximum capacity and the customer's demand.

Instead of reconstructing the weights, \citeauthor{carvalho2021} estimate the \textit{input shares} $\alpha_{ij}$ of each link,
$$
    \alpha_{ij} = \frac{\omega_{ij}}{\sum_i \omega_{ij}}.
$$ 
For any given customer-supplier pair of firms $\left(i, j\right)$ in the data, they assign $\alpha_{ij}$ proportionally to the input-output table entry corresponding to industries $i$ and $j$ belong to, i.e., $\alpha_{ij} \propto \tilde{\omega}_{a_i b_j}$, and renormalize them to ensure  $\sum_{i} \alpha_{ij} = 1$.

Real-world scenarios often present situations where it is unfeasible to find weights that align with aggregate observations. In \cite{welburn2020}, the authors design an inference strategy that aims to minimize the discrepancy between reconstructed and observed aggregate properties of the network. More specifically, the authors observe that, given a binary network $G$, it is not always possible to assign weights $\omega_{ij}$ that satisfy constraints $\sum_j \omega_{ij} = \tilde{\omega}_i^{\text{out}}$ and $\sum_{j}\omega_{ji} = \tilde{\omega}^{\text{in}}_i$. Take as an example a firm $i$ who supplies only a single firm $j$, and assume that $i$ is the only supplier of $j$. The aggregate constraints will only be satisfied if $i$'s sales match exactly $j$'s expenses, $\tilde{\omega}_i^{\text{out}} = \tilde{\omega}_j^{\text{in}}$, a condition not always respected in the data. The authors solve this issue by introducing a `residual node' $r$ to capture the portion of the economy that is not covered by the network $G$. This node accounts for all the firms that are not present in the data. They propose to find the set of weights $\omega_{ij}$ that minimize the loss $\mathcal{L} = \sum_{i} \omega_{i, r} + \sum_i\omega_{r, i}$, where $\omega_{ij}$ are subject to the usual constraints.

Finally, \citeauthor{hazan2019maximum} reconstructs the weights for a complete stock-flow consistent economy, with households, firms, banks, and flows of money in the form of consumption, firm-to-firm payments, wages, and interest payments. After reconstructing the network using maximum entropy methods (Sec.~\ref{subsub:maximum-entropy-binary}), stock-flow consistency makes it possible to write a linear system for the weights, which can be solved using Non-Negative Least Squares to avoid negative values. 

The performance of the methods reviewed in this subsection is unfortunately unknown, as information on the real weights was not available to the authors, who could not compare their reconstructions to the respective ground truths. However, in the future, researchers using these methods could partially validate their results by comparing them to the empirical regularities observed in \cite{bacilieri2023} for weight distributions and the relationships between in- and out-degrees and strengths.

\begin{table}[ht]
\centering
\footnotesize
\renewcommand{\arraystretch}{1.4}
\begin{tabular}{r>{\raggedright\arraybackslash}p{1.7cm}>{\raggedright\arraybackslash}p{22mm}>{\raggedright\arraybackslash}p{35mm}cc}
    \toprule
        & \textit{Coverage} & \textit{Dataset} & \textit{Inputs} & \footnotesize{\textit{Probabilistic}} & MaxEnt \\
    \midrule
    \citet{InoueTodo2019} & National, Japan & Tokyo Shoko Research & Firm sales, national IOTs &  & \\
    \citet{carvalho2021} & National, Japan & Tokyo Shoko Research & Firm sales, national IOTs &  & \\
    \citet{welburn2020} & National, US & S\&P Capital IQ, EDGAR. & Firm sales and inputs (COGS). & & \\
    \citet{hazan2019maximum} & National, Czech Republic & Full IOTs & Full IOTs & & \\
    \citet{bacilieri2023reconstructing} & National, International & Factset, Ecuador VAT & Firm sales, intermediate expenses, network density & X & X \\
    \citet{Ialongo2022} & National & Dutch banks' transaction data  & Firm sales, intermediate expenses by sector, network density (for calibration)  & X  & X \\
    \bottomrule
\end{tabular}
\caption{Overview of the papers that infer supply network weights.}
\label{tab:weights-reconstruction-resume}
\end{table}

\subsection{Maximum entropy for weights inference}
\label{subsec:max-ent-weights}

Another way of predicting weights given some aggregate trade information is to use the maximum entropy principle. The intuition behind this principle is computing a distribution that is \textit{maximally non-committal} with respect to unknown information \cite{jaynes1957}b or, in simpler words, to build a distribution that minimizes unjustified assumptions about the network. In Sec. \ref{subsub:maximum-entropy-binary}, we saw how maximum entropy can be used to compute probabilities for possible binary networks. We are now going to see how it can be used to predict weights.

If we consider the weights $\omega_{ij}$, subject to the (``hard'') constraints $\sum_j \omega_{ij} = \tilde{\omega}_i^{out}$, and $\sum_{j}\omega_{ji} = \tilde{\omega}^{in}_i$, where $\tilde{\omega}_i^{out}$ and $\tilde{\omega}_i^{in}$ represent the observed total outflow (intermediate sales) and inflow (intermediate expenses) of firm $i$, we find that the set of weights that maximize the Shannon Entropy 
$$
    \mathcal{S} = -\sum_i^N\sum_j^N \omega_{ij}\ln\omega_{ij},
$$
are
\begin{equation}
\label{eq:maxent-weights}
    \omega_{ij} = \frac{\tilde{\omega}^{out}_i\tilde{\omega}^{in}_j}{\tilde{\Omega}},
\end{equation}
where $\tilde{\Omega} = \sum_i \tilde{\omega}^{out}_i = \sum_i \tilde{\omega}^{in}_i$. This approach was also used in \cite{reisch2022inferring} for an undisclosed European country\footnote{
\citeauthor{bartolucci2023ranking} show that ``upstreamness'', a classic metric in I-O economics, can be recovered very well from networks reconstructed from maximum entropy, as long as the networks are not too sparse. This is because, under very general conditions for the original network, the first-order approximation of a node's upstreamness is its upstreamness in the maximum entropy-reconstructed network \cite{bartolucci2020}.
}.

A different application of the maximum-entropy principle, where constraints are imposed softly (see Sec.~\ref{subsub:link-prediction}), results in the solution used in \cite{bacilieri2023reconstructing} to reconstruct Ecuador's national production network and in \cite{Ialongo2022} to reconstruct the transaction network between customers of two Dutch banks. Building on \cite{parisi2020}, these papers first reconstruct the network's topology\footnote{In the case of \cite{bacilieri2023}, the topology is assumed to be known.}, then sample the (positive) weights $\omega_{ij}$ of the existing links from an exponential distribution,
$$
P\left(\omega_{ij} = x\right) = \beta_{ij}\exp\left(-\beta_{ij}x\right),
$$
where $\beta_{ij}$ is selected so that the expected value of $\omega_{ij}$, conditional to the existence of a link, is
$$
  \mathbb{E}_{ij}\left[\omega_{ij}| A_{ij} = 1\right] = \frac{\tilde{\omega}_i^{out}\tilde{\omega}_j^{in}}{p_{ij}\sum_i \tilde{\omega}^{out}_i}.
$$
In \cite{Ialongo2022}, $p_{ij}$ is defined by Eq.~\eqref{eq:p_ij-stripe-max-ent}. In contrast, \cite{bacilieri2023reconstructing} omits sector-specific constraints for intermediate inputs\footnote{
\cite{Ialongo2022} simply assume that the meso-level constraints are observable since they have this in their firm-level data. \cite{InoueTodo2019, hillman2021, carvalho2021} cannot read this information from the data, so they take meso-level information from the I-O tables. \cite{bacilieri2023reconstructing} argue that differences in accounting standards between firm- and industry-level networks are large so that the meso-level structure of a firm network should not be constrained to be like the I-O tables. \cite{bacilieri2023} shows that there are indeed some important differences, especially in industries that follow different accounting conventions, such as retail and wholesale trade.
}, and defines $p_{ij}$ as 
$$
p_{ij} = \frac{z\tilde{\omega}^{out}_i\tilde{\omega}_{j}^{in}}{1 + z\tilde{\omega}^{out}_i\tilde{\omega}_{j}^{in}}.
$$

Ref. \cite{bacilieri2023reconstructing} reports a cosine similarity of $0.928$ between inferred and actual weights, and also compute a few ``higher-order'' properties of the nodes that describe the propagation of shocks in production networks in an established macroeconomic model \cite{acemoglu2012network}, which the reconstructed network fails to capture adequately (the cosine similarity for the most relevant property, the \textit{influence vector}, is $\sim 0.5$).

In \cite{Ialongo2022}, visual inspection of the results shows a substantial enhancement in weight reconstruction when applying sector-specific constraints to firms' inputs, further underscoring the crucial role the economy's sectoral structure plays in the accurate reconstruction of production networks.

\section{Discussion}
\label{sec:discussion}

In this section, we take stock of what we can learn from existing studies, and provide suggestions on how the field could be further advanced.

\subsection{What have we learned?}

A first, clear message from the review is that in the context of supply networks, knowing the kind of product a firm makes is extremely important and substantially improves the reconstruction. This is evident both in the link prediction studies on industry data \cite{Brintrup2018}, commercial or country-level data \cite{mungo2023}, and in the maximum entropy reconstruction on payment data \cite{Ialongo2022}. Unsurprisingly, ongoing research tries to predict the firms' products at a granular level, for instance from websites \cite{occhini2023measuring}.

Second, the importance of products leads us to ask: to what extent can we, or should we rely on existing (national or inter-country) input-output matrices? While some studies reconstruct weights (conditional on links) using I-O links \cite{InoueTodo2019,carvalho2021,hillman2021}, others refrain from doing so \cite{bacilieri2023reconstructing}, by fear that differences in accounting conventions \cite{bacilieri2023} may create inconsistencies. Here the answer may depend on the goal of the reconstruction (see next section). A useful avenue for further research, however, would be to develop methods that easily make it possible to switch between business- and national accounting conventions. Such methods would necessarily use techniques and assumptions to allocate flows of money based on partially observed data, so that the methods reviewed here may be helpful.

Third, we have seen that more sophisticated machine learning methods do provide substantial boosts in performance. This is clear from the improvement in link prediction performance between the logistic regression and graph neural nets in the automotive dataset \cite{Brintrup2018, Kosasih2022}, and between simpler methods and gradient boosting in \citet{mungo2023} \footnote{However, in both studies, predictions made by sophisticated models are harder to interpret.}.

Fourth, there appears to be substantial scope for improving performance using ``alternative'' data. \citet{zhang_et_al_2012} and \citet{wichmann2020} have provided a proof of concept that mining news and websites for supplier-buyer relations can be automated, and we have already mentioned that websites can be an important source of key metadata for link prediction (especially product-related information). While phone data is likely to be difficult to access, it is worth remembering the impressive result in \cite{reisch2022inferring} that firms with average daily communication of more than $30s/$day have a 90\% probability of being connected.

A related question for further research will be to establish the potential of ``dynamical'' data. \citet{mungo2023revealing} showed that while there is information about the network in the sales growth rates correlation matrix, predicting the network remains difficult, as the distribution of pairwise correlation for connected and unconnected pairs overlaps greatly, even though their average is statistically significantly different. Nevertheless, there are interesting developments in this area for networks generally, with only one application to supply networks. One limitation has been that very few supply networks' datasets have a reasonable time-series dimension, but as these become more common it will perhaps become possible to find other firm-level dynamical features that contain fingerprints of their network.

Finally, many studies have shown that baking sensible economic intuition into the models usually improves predictions. To sum up, we have learned (or confirmed from existing literature) that link formation is likely driven by the kind of products firms make, their geographical distance, and their size. We have seen that firms who communicate a lot are likely to be in a supply-buy relationship and that firms that are in a relationship are likely to have a substantial co-movement in sales. While prediction is in some cases the ultimate goal, making methods that prioritize performance over interpretability appropriate \cite{Kosasih2022}, the quest for better reconstruction models has also prompted a deeper investigation into the behavioural and economic principles influencing how firms make and unmake their connections \cite{Brintrup2018, mungo2023}. Currently, no fully realistic supply network formation model has been developed (however, see \cite{AtalayEtAl2011} for an early example); we anticipate that reconstruction methods and the development of null models will, at least partly, go hand in hand.

\subsection{How can we learn more?}

What method works best for which task? We are not yet able to properly answer this question because the literature uses different datasets, takes different features of the data to make predictions, and uses different evaluation metrics. While this is warranted by the diversity of goals and applications, we think it would be valuable to organize ``horse races'', as has been done for financial networks \cite{anand2018}, and provide standard datasets, as is common in the machine learning community.

Let us first discuss the lack of comparability between studies. The methods proposed are very diverse and usually require distinct data to operate. The diversity of datasets and features used is understandable and valuable. For example, \citet{Kosasih2022} use topological features because one of their realistic use cases is to augment an existing ``observed'' network dataset, while \citet{mungo2023} avoid using topological information because their envisioned use case is to port a trained model to a context where no such features are available. As another example, while phone data is very hard to access, the study using this data made it possible to evaluate the systemic risk of each firm in an entire European country. 

A slightly less justified ``diversity of approaches'' is the lack of standardized assessment metrics, as it is in principle relatively easy to report several metrics.

Traditional statistical indicators (accuracy, AUROC, PR-AUC) provide an easy, well-known benchmark, and have already been functional in, e.g., propelling the development of computer-vision models \cite{imagenet2015}. Yet, the question remains as to whether they are sufficient to evaluate the reconstruction of a network, and what additional metrics should be adopted to supplement them. Some metrics, initially conceived for balanced datasets, may not hold up as reliably when applied to sparse networks, where non-existing links greatly outnumber the existing ones, further complicating the comparison between methods. Overall, the area under the Receiving Operator Characteristic Curve (AUROC) seems robust in the face of class imbalance: if one makes the imbalance more and more severe, its value does not change substantially (see Supplementary Material \cite{mungo2023}). Consequently, AUROC is a sensible metric to compare results. The area under the Precision-Recall curve (PR-AUC), which is more sensitive to the performance of the model on the minority class, is also very sensitive to the level of imbalance in the data; PR-AUC and imbalance should always be reported jointly.

Reporting basic topology metrics of the reconstructed network is also a sensible approach, as there is substantial evidence \cite{bacilieri2023} that some topological properties are universally shared by all production networks. For instance, \citet{bacilieri2023} showed that the tail exponents for the in- and out-degree distributions are remarkably similar in national, VAT-assembled datasets.

Ultimately, as we plug reconstructed networks into economic models, the optimal metric will be the one that best correlates with accurate economic predictions. Identifying these proper ``dynamical'' indicators needs to go hand-in-hand with the development of economic models that are carefully validated on real-world data and can become legitimate standards for evaluating reconstruction performance.

While agreeing on a set of metrics and features appears relatively easy, the key challenge ahead is data availability. To follow our previous analogy, in computer vision, researchers can access standard, large-scale datasets \cite{imagenet2009} of annotated images to train and evaluate their models. Similar datasets for production network reconstruction are not currently available and, due to the confidential or proprietary nature of such data, its assembly seems unlikely in the near future. The research community should unite to devise strategies to circumvent this issue, possibly by considering the use of synthetic data \cite{synthetic2022} as an alternative to real data. While synthetic data generation is currently an active and exciting area of research, it is less well-developed for networks than for tabular data and still suffers from either a lack of privacy guarantees (for traditional methods) or a lack of interpretability of the privacy guarantees (for differential privacy).

\section{Two research agendas}
\label{sec:agendas}

For many practical applications, it is necessary to know much more than the value of transactions between firms. We lay out two research programs \-- one that aims to reconstruct supply networks to allow for real-time monitoring of disruptions and logistics optimization; and one that aims to reconstruct a granular version of global macroeconomic datasets.

\subsection{Towards supply chain visibility for risk management}

Past decades have been characterized by supply chain cost optimization objectives, which have led to just-in-time initiatives that stripped buffer inventories from supply lines, that have already become geographically longer with offshoring practices. 

While high-impact, rare events such as COVID-19 highlighted the vulnerability of these global, highly complex modes of operation, organisations often struggle with increased volatility in their day-to-day procurement.  Supply chain researchers are increasingly seeking methods to build resilience in what is now frequently termed a “shortage economy” \cite{ivanov2022shortage}. However, these efforts are often hindered by a lack of visibility on supply chain dependencies as companies do not disclose commercially sensitive information such as whom they buy goods and services from.  

As link prediction and reconstruction methods presented in this paper do not rely on companies’ willingness to share data, they have the potential to become a primary toolset in supply chain risk management. Our review shows that buyer-supplier link prediction is possible with various differing methodologies and success rates. Recently proposed methods for reconstructing knowledge graphs go beyond who-supplies-whom, but also enable prediction of other types of relevant information such as where firms are located, and what they produce, paving the way for a new era of ``digital supply chain surveillance'' \cite{brintrup2022digital}. 

Much further work is needed in this context. For instance, use cases that evaluate how the identification of risky supplier locations and production dependencies might help with effective mitigation strategies such as multi-sourcing, supply chain reconfiguration, insurance, or inventory buffers. Beyond addressing supply disruption risk, an understanding of supply chain structure could be informative for the detection of supply chain fraud and counterfeit products. Improved visibility may help improve regulatory compliance on the Environmental, Social and Governance (ESG) practice. Methods that help detect transaction volumes could improve supply chain financing in which lenders often struggle with identifying financial risk exposure. To achieve these, different ontologies are needed to be built and integrated into existing knowledge graph completion methods. New methods for inferring compliance, fraud, and other areas of interest from knowledge graphs need to be developed. Lastly, any resulting graph will be limited by underlying assumptions and incomplete data, which, in turn, may be shaped by the observable data at hand. Hence data imputation and uncertainty quantification will need to inform the resulting graphs.

\subsection{Towards granular global economic accounts}

For macroeconomic applications, our interest extends beyond the mere flow of money between firms. Macroeconomics concerns quantities such as GDP, which represents at the same time the total income, total expenditures, and total ``value added'' of an economy. Firm-to-firm transactions are not sufficient to truly understand how economic agents create value, redistribute income, and spend on goods and services. 

As a result, to support the development of large-scale realistic agent-based models, we need an ambitious agenda to develop semi-synthetic populations, which would include all the available micro information and supplement it by synthetic micro data in a way that results in meso- and macro-level aggregates compatible with observed meso- and macro-level quantities typically observable from national accounts. We elaborate briefly on three strands of research within this agenda.

First, it will be important to ensure compatibility between micro and meso-level data, which are usually compiled using different accounting rules. National accounting principles provide a solid conceptual framework, so developing reconstructed datasets that respect these principles would have many advantages, including making it easier to use this data in macro models, easier to improve this data using macro-level information, and easier to match this data with other relevant datasets. However, firm-level data is usually compiled using business accounting rules, so that simply ``summing up'' firm-level data does not necessarily lead to the supposedly equivalent concept in national accounts. As we have highlighted, this is a potential to, for instance, use IOTs as additional information to reconstruct firm-level networks.

Second, modern work in economics shows that employer-employee-matched data and datasets on consumer baskets' heterogeneity are crucial to understanding inequality, long-run growth, or carbon emissions. As a result, a straightforward extension of the ``reconstruction of economic networks'' program would be to predict employer-employee relations and consumer-firm relations (See \cite{hazan2019maximum} for a first attempt). Existing efforts to develop data-driven agent-based models rely on such synthetic populations. While there exists a lot of work on recommender systems for suggesting products to consumers, and more recently some work on career suggestions, these efforts have not been leveraged to create reliable synthetic populations.

Third, many of the studies presented here worked with money flows, omitting a distinction between prices and quantities. This is driven by the fact that firm-level supply networks with both price and quantity information are very rare, but this is a serious issue for economic modelling, where prices obviously play a key role. To model inflation, understand growth and business cycles, we need measures of quantities produced (or inflation-adjusted values). New methods for inferring prices, perhaps based on companies' websites and other features, would be very extremely helpful in this context.

\section{Conclusion}

The reconstruction of supply networks through mathematical methods is a young field. This paper offers a review of methodologies that researchers have proposed to grapple with this challenge.

Good proof-of-concept studies exist, but much remains to be done. A striking feature of the literature is the diversity of methods, datasets and evaluation metrics. While this is justified by the different backgrounds and motivations of the researchers, we think that progress in this area would benefit from the availability of open datasets and the definition of standard metrics, so that horse races could be organised. 

We were able to propose some guidelines to standardize performance metrics, but the path to open datasets is more complicated and will require international cooperation that either facilitates researchers' access, or fosters the creation of high-fidelity synthetic datasets.

Despite this difficulty, we think that reconstructing supply networks is an excellent playing ground for the complex systems community, as it requires a deep understanding of networks, statistics, and dynamical systems, together with an appreciation that these networks emerge from the decentralized interactions of millions of highly heterogenous, bounded-rational agents operating with different objectives at different time scales.

\bibliography{biblio_clean}

\begin{thebibliography}{61}%
\makeatletter
\providecommand \@ifxundefined [1]{%
 \@ifx{#1\undefined}
}%
\providecommand \@ifnum [1]{%
 \ifnum #1\expandafter \@firstoftwo
 \else \expandafter \@secondoftwo
 \fi
}%
\providecommand \@ifx [1]{%
 \ifx #1\expandafter \@firstoftwo
 \else \expandafter \@secondoftwo
 \fi
}%
\providecommand \natexlab [1]{#1}%
\providecommand \enquote  [1]{``#1''}%
\providecommand \bibnamefont  [1]{#1}%
\providecommand \bibfnamefont [1]{#1}%
\providecommand \citenamefont [1]{#1}%
\providecommand \href@noop [0]{\@secondoftwo}%
\providecommand \href [0]{\begingroup \@sanitize@url \@href}%
\providecommand \@href[1]{\@@startlink{#1}\@@href}%
\providecommand \@@href[1]{\endgroup#1\@@endlink}%
\providecommand \@sanitize@url [0]{\catcode `\\12\catcode `\$12\catcode `\&12\catcode `\#12\catcode `\^12\catcode `\_12\catcode `\%12\relax}%
\providecommand \@@startlink[1]{}%
\providecommand \@@endlink[0]{}%
\providecommand \url  [0]{\begingroup\@sanitize@url \@url }%
\providecommand \@url [1]{\endgroup\@href {#1}{\urlprefix }}%
\providecommand \urlprefix  [0]{URL }%
\providecommand \Eprint [0]{\href }%
\providecommand \doibase [0]{https://doi.org/}%
\providecommand \selectlanguage [0]{\@gobble}%
\providecommand \bibinfo  [0]{\@secondoftwo}%
\providecommand \bibfield  [0]{\@secondoftwo}%
\providecommand \translation [1]{[#1]}%
\providecommand \BibitemOpen [0]{}%
\providecommand \bibitemStop [0]{}%
\providecommand \bibitemNoStop [0]{.\EOS\space}%
\providecommand \EOS [0]{\spacefactor3000\relax}%
\providecommand \BibitemShut  [1]{\csname bibitem#1\endcsname}%
\let\auto@bib@innerbib\@empty
\bibitem [{\citenamefont {Haldane}\ and\ \citenamefont {May}(2011)}]{haldane_systemic_2011}%
  \BibitemOpen
  \bibfield  {author} {\bibinfo {author} {\bibfnamefont {A.~G.}\ \bibnamefont {Haldane}}\ and\ \bibinfo {author} {\bibfnamefont {R.~M.}\ \bibnamefont {May}},\ }\bibfield  {title} {\bibinfo {title} {Systemic risk in banking ecosystems},\ }\href {https://doi.org/10.1038/nature09659} {\bibfield  {journal} {\bibinfo  {journal} {Nature}\ }\textbf {\bibinfo {volume} {469}},\ \bibinfo {pages} {351} (\bibinfo {year} {2011})}\BibitemShut {NoStop}%
\bibitem [{\citenamefont {Bardoscia}\ \emph {et~al.}(2021)\citenamefont {Bardoscia}, \citenamefont {Barucca}, \citenamefont {Battiston}, \citenamefont {Caccioli}, \citenamefont {Cimini}, \citenamefont {Garlaschelli}, \citenamefont {Saracco}, \citenamefont {Squartini},\ and\ \citenamefont {Caldarelli}}]{bardoscia_physics_2021}%
  \BibitemOpen
  \bibfield  {author} {\bibinfo {author} {\bibfnamefont {M.}~\bibnamefont {Bardoscia}}, \bibinfo {author} {\bibfnamefont {P.}~\bibnamefont {Barucca}}, \bibinfo {author} {\bibfnamefont {S.}~\bibnamefont {Battiston}}, \bibinfo {author} {\bibfnamefont {F.}~\bibnamefont {Caccioli}}, \bibinfo {author} {\bibfnamefont {G.}~\bibnamefont {Cimini}}, \bibinfo {author} {\bibfnamefont {D.}~\bibnamefont {Garlaschelli}}, \bibinfo {author} {\bibfnamefont {F.}~\bibnamefont {Saracco}}, \bibinfo {author} {\bibfnamefont {T.}~\bibnamefont {Squartini}},\ and\ \bibinfo {author} {\bibfnamefont {G.}~\bibnamefont {Caldarelli}},\ }\bibfield  {title} {\bibinfo {title} {The physics of financial networks},\ }\href {https://doi.org/10.1038/s42254-021-00322-5} {\bibfield  {journal} {\bibinfo  {journal} {Nature Reviews Physics}\ }\textbf {\bibinfo {volume} {3}},\ \bibinfo {pages} {490} (\bibinfo {year} {2021})}\BibitemShut {NoStop}%
\bibitem [{\citenamefont {Bacilieri}\ \emph {et~al.}(2023)\citenamefont {Bacilieri}, \citenamefont {Borsos}, \citenamefont {Astudillo-Estevez},\ and\ \citenamefont {Lafond}}]{bacilieri2023}%
  \BibitemOpen
  \bibfield  {author} {\bibinfo {author} {\bibfnamefont {A.}~\bibnamefont {Bacilieri}}, \bibinfo {author} {\bibfnamefont {A.}~\bibnamefont {Borsos}}, \bibinfo {author} {\bibfnamefont {P.}~\bibnamefont {Astudillo-Estevez}},\ and\ \bibinfo {author} {\bibfnamefont {F.}~\bibnamefont {Lafond}},\ }\href@noop {} {\emph {\bibinfo {title} {Firm-level production networks: what do we (really) know?}}},\ \bibinfo {type} {Tech. Rep.}\ \bibinfo {number} {2023-08}\ (\bibinfo  {institution} {Institute for New Economic Thinking},\ \bibinfo {year} {2023})\BibitemShut {NoStop}%
\bibitem [{\citenamefont {Newman}(2002)}]{newman2022}%
  \BibitemOpen
  \bibfield  {author} {\bibinfo {author} {\bibfnamefont {M.~E.~J.}\ \bibnamefont {Newman}},\ }\bibfield  {title} {\bibinfo {title} {Assortative mixing in networks},\ }\href {https://doi.org/10.1103/PhysRevLett.89.208701} {\bibfield  {journal} {\bibinfo  {journal} {Phys. Rev. Lett.}\ }\textbf {\bibinfo {volume} {89}},\ \bibinfo {pages} {208701} (\bibinfo {year} {2002})}\BibitemShut {NoStop}%
\bibitem [{\citenamefont {Demir}\ \emph {et~al.}(0)\citenamefont {Demir}, \citenamefont {Fieler}, \citenamefont {Xu},\ and\ \citenamefont {Yang}}]{demir2023}%
  \BibitemOpen
  \bibfield  {author} {\bibinfo {author} {\bibfnamefont {B.}~\bibnamefont {Demir}}, \bibinfo {author} {\bibfnamefont {A.~C.}\ \bibnamefont {Fieler}}, \bibinfo {author} {\bibfnamefont {D.~Y.}\ \bibnamefont {Xu}},\ and\ \bibinfo {author} {\bibfnamefont {K.~K.}\ \bibnamefont {Yang}},\ }\bibfield  {title} {\bibinfo {title} {O-ring production networks},\ }\href {https://doi.org/10.1086/725703} {\bibfield  {journal} {\bibinfo  {journal} {Journal of Political Economy}\ }\textbf {\bibinfo {volume} {0}},\ \bibinfo {pages} {null} (\bibinfo {year} {0})},\ \Eprint {https://arxiv.org/abs/https://doi.org/10.1086/725703} {https://doi.org/10.1086/725703} \BibitemShut {NoStop}%
\bibitem [{\citenamefont {Bernard}\ \emph {et~al.}(2019)\citenamefont {Bernard}, \citenamefont {Moxnes},\ and\ \citenamefont {Saito}}]{bernard2019b}%
  \BibitemOpen
  \bibfield  {author} {\bibinfo {author} {\bibfnamefont {A.~B.}\ \bibnamefont {Bernard}}, \bibinfo {author} {\bibfnamefont {A.}~\bibnamefont {Moxnes}},\ and\ \bibinfo {author} {\bibfnamefont {Y.~U.}\ \bibnamefont {Saito}},\ }\bibfield  {title} {\bibinfo {title} {Production networks, geography, and firm performance},\ }\href {https://doi.org/10.1086/700764} {\bibfield  {journal} {\bibinfo  {journal} {Journal of Political Economy}\ }\textbf {\bibinfo {volume} {127}},\ \bibinfo {pages} {639} (\bibinfo {year} {2019})},\ \Eprint {https://arxiv.org/abs/https://doi.org/10.1086/700764} {https://doi.org/10.1086/700764} \BibitemShut {NoStop}%
\bibitem [{\citenamefont {Christopher}\ and\ \citenamefont {Holweg}(2011)}]{christopher2011supply}%
  \BibitemOpen
  \bibfield  {author} {\bibinfo {author} {\bibfnamefont {M.}~\bibnamefont {Christopher}}\ and\ \bibinfo {author} {\bibfnamefont {M.}~\bibnamefont {Holweg}},\ }\bibfield  {title} {\bibinfo {title} {“supply chain 2.0”: Managing supply chains in the era of turbulence},\ }\href@noop {} {\bibfield  {journal} {\bibinfo  {journal} {International journal of physical distribution \& logistics management}\ }\textbf {\bibinfo {volume} {41}},\ \bibinfo {pages} {63} (\bibinfo {year} {2011})}\BibitemShut {NoStop}%
\bibitem [{\citenamefont {Choi}\ and\ \citenamefont {Hong}(2002)}]{choi2002unveiling}%
  \BibitemOpen
  \bibfield  {author} {\bibinfo {author} {\bibfnamefont {T.~Y.}\ \bibnamefont {Choi}}\ and\ \bibinfo {author} {\bibfnamefont {Y.}~\bibnamefont {Hong}},\ }\bibfield  {title} {\bibinfo {title} {Unveiling the structure of supply networks: case studies in {H}onda, {A}cura, and {D}aimler{C}hrysler},\ }\href@noop {} {\bibfield  {journal} {\bibinfo  {journal} {Journal of Operations Management}\ }\textbf {\bibinfo {volume} {20}},\ \bibinfo {pages} {469} (\bibinfo {year} {2002})}\BibitemShut {NoStop}%
\bibitem [{\citenamefont {Brintrup}\ \emph {et~al.}(2016)\citenamefont {Brintrup}, \citenamefont {Ledwoch},\ and\ \citenamefont {Barros}}]{brintrup2016topological}%
  \BibitemOpen
  \bibfield  {author} {\bibinfo {author} {\bibfnamefont {A.}~\bibnamefont {Brintrup}}, \bibinfo {author} {\bibfnamefont {A.}~\bibnamefont {Ledwoch}},\ and\ \bibinfo {author} {\bibfnamefont {J.}~\bibnamefont {Barros}},\ }\bibfield  {title} {\bibinfo {title} {Topological robustness of the global automotive industry},\ }\href@noop {} {\bibfield  {journal} {\bibinfo  {journal} {Logistics Research}\ }\textbf {\bibinfo {volume} {9}},\ \bibinfo {pages} {1} (\bibinfo {year} {2016})}\BibitemShut {NoStop}%
\bibitem [{\citenamefont {Brintrup}\ \emph {et~al.}(2015)\citenamefont {Brintrup}, \citenamefont {Wang},\ and\ \citenamefont {Tiwari}}]{brintrup2015supply}%
  \BibitemOpen
  \bibfield  {author} {\bibinfo {author} {\bibfnamefont {A.}~\bibnamefont {Brintrup}}, \bibinfo {author} {\bibfnamefont {Y.}~\bibnamefont {Wang}},\ and\ \bibinfo {author} {\bibfnamefont {A.}~\bibnamefont {Tiwari}},\ }\bibfield  {title} {\bibinfo {title} {Supply networks as complex systems: a network-science-based characterization},\ }\href@noop {} {\bibfield  {journal} {\bibinfo  {journal} {IEEE Systems Journal}\ }\textbf {\bibinfo {volume} {11}},\ \bibinfo {pages} {2170} (\bibinfo {year} {2015})}\BibitemShut {NoStop}%
\bibitem [{\citenamefont {Perera}\ \emph {et~al.}(2017)\citenamefont {Perera}, \citenamefont {Bell},\ and\ \citenamefont {Bliemer}}]{perera2017network}%
  \BibitemOpen
  \bibfield  {author} {\bibinfo {author} {\bibfnamefont {S.}~\bibnamefont {Perera}}, \bibinfo {author} {\bibfnamefont {M.~G.}\ \bibnamefont {Bell}},\ and\ \bibinfo {author} {\bibfnamefont {M.~C.}\ \bibnamefont {Bliemer}},\ }\bibfield  {title} {\bibinfo {title} {Network science approach to modelling the topology and robustness of supply chain networks: a review and perspective},\ }\href@noop {} {\bibfield  {journal} {\bibinfo  {journal} {Applied network science}\ }\textbf {\bibinfo {volume} {2}},\ \bibinfo {pages} {1} (\bibinfo {year} {2017})}\BibitemShut {NoStop}%
\bibitem [{\citenamefont {Wichmann}\ \emph {et~al.}(2020)\citenamefont {Wichmann}, \citenamefont {Brintrup}, \citenamefont {Baker}, \citenamefont {Woodall},\ and\ \citenamefont {McFarlane}}]{wichmann2020}%
  \BibitemOpen
  \bibfield  {author} {\bibinfo {author} {\bibfnamefont {P.}~\bibnamefont {Wichmann}}, \bibinfo {author} {\bibfnamefont {A.}~\bibnamefont {Brintrup}}, \bibinfo {author} {\bibfnamefont {S.}~\bibnamefont {Baker}}, \bibinfo {author} {\bibfnamefont {P.}~\bibnamefont {Woodall}},\ and\ \bibinfo {author} {\bibfnamefont {D.}~\bibnamefont {McFarlane}},\ }\bibfield  {title} {\bibinfo {title} {Extracting supply chain maps from news articles using deep neural networks},\ }\href {https://doi.org/10.1080/00207543.2020.1720925} {\bibfield  {journal} {\bibinfo  {journal} {International Journal of Production Research}\ }\textbf {\bibinfo {volume} {58}},\ \bibinfo {pages} {5320} (\bibinfo {year} {2020})},\ \Eprint {https://arxiv.org/abs/https://doi.org/10.1080/00207543.2020.1720925} {https://doi.org/10.1080/00207543.2020.1720925} \BibitemShut {NoStop}%
\bibitem [{\citenamefont {Ialongo}\ \emph {et~al.}(2022)\citenamefont {Ialongo}, \citenamefont {de~Valk}, \citenamefont {Marchese}, \citenamefont {Jansen}, \citenamefont {Zmarrou}, \citenamefont {Squartini},\ and\ \citenamefont {Garlaschelli}}]{Ialongo2022}%
  \BibitemOpen
  \bibfield  {author} {\bibinfo {author} {\bibfnamefont {L.~N.}\ \bibnamefont {Ialongo}}, \bibinfo {author} {\bibfnamefont {C.}~\bibnamefont {de~Valk}}, \bibinfo {author} {\bibfnamefont {E.}~\bibnamefont {Marchese}}, \bibinfo {author} {\bibfnamefont {F.}~\bibnamefont {Jansen}}, \bibinfo {author} {\bibfnamefont {H.}~\bibnamefont {Zmarrou}}, \bibinfo {author} {\bibfnamefont {T.}~\bibnamefont {Squartini}},\ and\ \bibinfo {author} {\bibfnamefont {D.}~\bibnamefont {Garlaschelli}},\ }\bibfield  {title} {\bibinfo {title} {Reconstructing firm-level interactions in the dutch input–output network from production constraints},\ }\href {https://doi.org/10.1038/s41598-022-13996-3} {\bibfield  {journal} {\bibinfo  {journal} {Scientific Reports}\ }\textbf {\bibinfo {volume} {12}},\ \bibinfo {pages} {11847} (\bibinfo {year} {2022})},\ \bibinfo {note} {number: 1 Publisher: Nature Publishing Group}\BibitemShut {NoStop}%
\bibitem [{\citenamefont {Brintrup}\ \emph {et~al.}(2018)\citenamefont {Brintrup}, \citenamefont {Wichmann}, \citenamefont {Woodall}, \citenamefont {McFarlane}, \citenamefont {Nicks},\ and\ \citenamefont {Krechel}}]{Brintrup2018}%
  \BibitemOpen
  \bibfield  {author} {\bibinfo {author} {\bibfnamefont {A.}~\bibnamefont {Brintrup}}, \bibinfo {author} {\bibfnamefont {P.}~\bibnamefont {Wichmann}}, \bibinfo {author} {\bibfnamefont {P.}~\bibnamefont {Woodall}}, \bibinfo {author} {\bibfnamefont {D.}~\bibnamefont {McFarlane}}, \bibinfo {author} {\bibfnamefont {E.}~\bibnamefont {Nicks}},\ and\ \bibinfo {author} {\bibfnamefont {W.}~\bibnamefont {Krechel}},\ }\bibfield  {title} {\bibinfo {title} {Predicting hidden links in supply networks},\ }\href {https://doi.org/10.1155/2018/9104387} {\bibfield  {journal} {\bibinfo  {journal} {Complexity}\ }\textbf {\bibinfo {volume} {2018}},\ \bibinfo {pages} {9104387} (\bibinfo {year} {2018})}\BibitemShut {NoStop}%
\bibitem [{\citenamefont {Lü}\ and\ \citenamefont {Zhou}(2011)}]{lu2011}%
  \BibitemOpen
  \bibfield  {author} {\bibinfo {author} {\bibfnamefont {L.}~\bibnamefont {Lü}}\ and\ \bibinfo {author} {\bibfnamefont {T.}~\bibnamefont {Zhou}},\ }\bibfield  {title} {\bibinfo {title} {Link prediction in complex networks: A survey},\ }\href {https://doi.org/https://doi.org/10.1016/j.physa.2010.11.027} {\bibfield  {journal} {\bibinfo  {journal} {Physica A: Statistical Mechanics and its Applications}\ }\textbf {\bibinfo {volume} {390}},\ \bibinfo {pages} {1150} (\bibinfo {year} {2011})}\BibitemShut {NoStop}%
\bibitem [{\citenamefont {Squartini}\ \emph {et~al.}(2018)\citenamefont {Squartini}, \citenamefont {Caldarelli}, \citenamefont {Cimini}, \citenamefont {Gabrielli},\ and\ \citenamefont {Garlaschelli}}]{SquartiniEtAl2018}%
  \BibitemOpen
  \bibfield  {author} {\bibinfo {author} {\bibfnamefont {T.}~\bibnamefont {Squartini}}, \bibinfo {author} {\bibfnamefont {G.}~\bibnamefont {Caldarelli}}, \bibinfo {author} {\bibfnamefont {G.}~\bibnamefont {Cimini}}, \bibinfo {author} {\bibfnamefont {A.}~\bibnamefont {Gabrielli}},\ and\ \bibinfo {author} {\bibfnamefont {D.}~\bibnamefont {Garlaschelli}},\ }\bibfield  {title} {\bibinfo {title} {Reconstruction methods for networks: The case of economic and financial systems},\ }\href {https://doi.org/https://doi.org/10.1016/j.physrep.2018.06.008} {\bibfield  {journal} {\bibinfo  {journal} {Physics Reports}\ }\textbf {\bibinfo {volume} {757}},\ \bibinfo {pages} {1} (\bibinfo {year} {2018})}\BibitemShut {NoStop}%
\bibitem [{\citenamefont {Diem}\ \emph {et~al.}(2022)\citenamefont {Diem}, \citenamefont {Borsos}, \citenamefont {Reisch}, \citenamefont {Kertész},\ and\ \citenamefont {Thurner}}]{diem2022quantifying}%
  \BibitemOpen
  \bibfield  {author} {\bibinfo {author} {\bibfnamefont {C.}~\bibnamefont {Diem}}, \bibinfo {author} {\bibfnamefont {A.}~\bibnamefont {Borsos}}, \bibinfo {author} {\bibfnamefont {T.}~\bibnamefont {Reisch}}, \bibinfo {author} {\bibfnamefont {J.}~\bibnamefont {Kertész}},\ and\ \bibinfo {author} {\bibfnamefont {S.}~\bibnamefont {Thurner}},\ }\bibfield  {title} {\bibinfo {title} {Quantifying firm-level economic systemic risk from nation-wide supply networks},\ }\href {https://doi.org/10.1038/s41598-022-11522-z} {\bibfield  {journal} {\bibinfo  {journal} {Scientific Reports}\ }\textbf {\bibinfo {volume} {12}},\ \bibinfo {pages} {7719} (\bibinfo {year} {2022})}\BibitemShut {NoStop}%
\bibitem [{\citenamefont {Mori}\ \emph {et~al.}(2012)\citenamefont {Mori}, \citenamefont {Kajikawa}, \citenamefont {Kashima},\ and\ \citenamefont {Sakata}}]{mori2012}%
  \BibitemOpen
  \bibfield  {author} {\bibinfo {author} {\bibfnamefont {J.}~\bibnamefont {Mori}}, \bibinfo {author} {\bibfnamefont {Y.}~\bibnamefont {Kajikawa}}, \bibinfo {author} {\bibfnamefont {H.}~\bibnamefont {Kashima}},\ and\ \bibinfo {author} {\bibfnamefont {I.}~\bibnamefont {Sakata}},\ }\bibfield  {title} {\bibinfo {title} {Machine learning approach for finding business partners and building reciprocal relationships},\ }\href {https://doi.org/https://doi.org/10.1016/j.eswa.2012.01.202} {\bibfield  {journal} {\bibinfo  {journal} {Expert Systems with Applications}\ }\textbf {\bibinfo {volume} {39}},\ \bibinfo {pages} {10402} (\bibinfo {year} {2012})}\BibitemShut {NoStop}%
\bibitem [{\citenamefont {Zuo}\ \emph {et~al.}(2016)\citenamefont {Zuo}, \citenamefont {Kajikawa},\ and\ \citenamefont {Mori}}]{zuo_kajikawa_mori2016}%
  \BibitemOpen
  \bibfield  {author} {\bibinfo {author} {\bibfnamefont {Y.}~\bibnamefont {Zuo}}, \bibinfo {author} {\bibfnamefont {Y.}~\bibnamefont {Kajikawa}},\ and\ \bibinfo {author} {\bibfnamefont {J.}~\bibnamefont {Mori}},\ }\bibfield  {title} {\bibinfo {title} {Extraction of business relationships in supply networks using statistical learning theory},\ }\href {https://doi.org/https://doi.org/10.1016/j.heliyon.2016.e00123} {\bibfield  {journal} {\bibinfo  {journal} {Heliyon}\ }\textbf {\bibinfo {volume} {2}},\ \bibinfo {pages} {e00123} (\bibinfo {year} {2016})}\BibitemShut {NoStop}%
\bibitem [{\citenamefont {Sasaki}\ and\ \citenamefont {Sakata}(2017)}]{sasaki_sakata_2017}%
  \BibitemOpen
  \bibfield  {author} {\bibinfo {author} {\bibfnamefont {H.}~\bibnamefont {Sasaki}}\ and\ \bibinfo {author} {\bibfnamefont {I.}~\bibnamefont {Sakata}},\ }\bibfield  {title} {\bibinfo {title} {Prediction of business partners using an n-gram-based approach that combines a network model and linear model of a supply chain},\ }in\ \href {https://doi.org/10.23919/PICMET.2017.8125410} {\emph {\bibinfo {booktitle} {2017 Portland International Conference on Management of Engineering and Technology (PICMET)}}}\ (\bibinfo {year} {2017})\ pp.\ \bibinfo {pages} {1--8}\BibitemShut {NoStop}%
\bibitem [{\citenamefont {Lee}\ and\ \citenamefont {Kim}(2022)}]{lee2022}%
  \BibitemOpen
  \bibfield  {author} {\bibinfo {author} {\bibfnamefont {D.}~\bibnamefont {Lee}}\ and\ \bibinfo {author} {\bibfnamefont {K.}~\bibnamefont {Kim}},\ }\bibfield  {title} {\bibinfo {title} {Business transaction recommendation for discovering potential business partners using deep learning},\ }\href {https://doi.org/https://doi.org/10.1016/j.eswa.2022.117222} {\bibfield  {journal} {\bibinfo  {journal} {Expert Systems with Applications}\ }\textbf {\bibinfo {volume} {201}},\ \bibinfo {pages} {117222} (\bibinfo {year} {2022})}\BibitemShut {NoStop}%
\bibitem [{\citenamefont {Kosasih}\ and\ \citenamefont {Brintrup}(2022)}]{Kosasih2022}%
  \BibitemOpen
  \bibfield  {author} {\bibinfo {author} {\bibfnamefont {E.~E.}\ \bibnamefont {Kosasih}}\ and\ \bibinfo {author} {\bibfnamefont {A.}~\bibnamefont {Brintrup}},\ }\bibfield  {title} {\bibinfo {title} {A machine learning approach for predicting hidden links in supply chain with graph neural networks},\ }\href {https://doi.org/10.1080/00207543.2021.1956697} {\bibfield  {journal} {\bibinfo  {journal} {International Journal of Production Research}\ }\textbf {\bibinfo {volume} {60}},\ \bibinfo {pages} {5380} (\bibinfo {year} {2022})},\ \Eprint {https://arxiv.org/abs/https://doi.org/10.1080/00207543.2021.1956697} {https://doi.org/10.1080/00207543.2021.1956697} \BibitemShut {NoStop}%
\bibitem [{\citenamefont {Minakawa}\ \emph {et~al.}(2023)\citenamefont {Minakawa}, \citenamefont {Izumi}, \citenamefont {Sakaji},\ and\ \citenamefont {Sano}}]{minakawa_et_al_2023}%
  \BibitemOpen
  \bibfield  {author} {\bibinfo {author} {\bibfnamefont {N.}~\bibnamefont {Minakawa}}, \bibinfo {author} {\bibfnamefont {K.}~\bibnamefont {Izumi}}, \bibinfo {author} {\bibfnamefont {H.}~\bibnamefont {Sakaji}},\ and\ \bibinfo {author} {\bibfnamefont {H.}~\bibnamefont {Sano}},\ }\bibfield  {title} {\bibinfo {title} {Transaction prediction by using graph neural network and textual industry information},\ }in\ \href@noop {} {\emph {\bibinfo {booktitle} {New Frontiers in Artificial Intelligence}}},\ \bibinfo {editor} {edited by\ \bibinfo {editor} {\bibfnamefont {Y.}~\bibnamefont {Takama}}, \bibinfo {editor} {\bibfnamefont {K.}~\bibnamefont {Yada}}, \bibinfo {editor} {\bibfnamefont {K.}~\bibnamefont {Satoh}},\ and\ \bibinfo {editor} {\bibfnamefont {S.}~\bibnamefont {Arai}}}\ (\bibinfo  {publisher} {Springer Nature Switzerland},\ \bibinfo {address} {Cham},\ \bibinfo {year} {2023})\ pp.\ \bibinfo {pages} {251--266}\BibitemShut {NoStop}%
\bibitem [{\citenamefont {Mungo}\ \emph {et~al.}(2023)\citenamefont {Mungo}, \citenamefont {Lafond}, \citenamefont {Astudillo-Estévez},\ and\ \citenamefont {{Doyne Farmer}}}]{mungo2023}%
  \BibitemOpen
  \bibfield  {author} {\bibinfo {author} {\bibfnamefont {L.}~\bibnamefont {Mungo}}, \bibinfo {author} {\bibfnamefont {F.}~\bibnamefont {Lafond}}, \bibinfo {author} {\bibfnamefont {P.}~\bibnamefont {Astudillo-Estévez}},\ and\ \bibinfo {author} {\bibfnamefont {J.}~\bibnamefont {{Doyne Farmer}}},\ }\bibfield  {title} {\bibinfo {title} {Reconstructing production networks using machine learning},\ }\href {https://doi.org/https://doi.org/10.1016/j.jedc.2023.104607} {\bibfield  {journal} {\bibinfo  {journal} {Journal of Economic Dynamics and Control}\ ,\ \bibinfo {pages} {104607}} (\bibinfo {year} {2023})}\BibitemShut {NoStop}%
\bibitem [{\citenamefont {Zhang}\ \emph {et~al.}(2012)\citenamefont {Zhang}, \citenamefont {Lau}, \citenamefont {Xia}, \citenamefont {Li},\ and\ \citenamefont {Li}}]{zhang_et_al_2012}%
  \BibitemOpen
  \bibfield  {author} {\bibinfo {author} {\bibfnamefont {W.}~\bibnamefont {Zhang}}, \bibinfo {author} {\bibfnamefont {R.~Y.}\ \bibnamefont {Lau}}, \bibinfo {author} {\bibfnamefont {Y.}~\bibnamefont {Xia}}, \bibinfo {author} {\bibfnamefont {C.}~\bibnamefont {Li}},\ and\ \bibinfo {author} {\bibfnamefont {W.~M.}\ \bibnamefont {Li}},\ }\bibfield  {title} {\bibinfo {title} {Latent business networks mining: A probabilistic generative model},\ }in\ \href {https://doi.org/10.1109/WI-IAT.2012.195} {\emph {\bibinfo {booktitle} {2012 IEEE/WIC/ACM International Conferences on Web Intelligence and Intelligent Agent Technology}}},\ Vol.~\bibinfo {volume} {1}\ (\bibinfo {year} {2012})\ pp.\ \bibinfo {pages} {558--562}\BibitemShut {NoStop}%
\bibitem [{\citenamefont {Schaffer~P.}(2023)}]{bert2023}%
  \BibitemOpen
  \bibfield  {author} {\bibinfo {author} {\bibfnamefont {B.~A.}\ \bibnamefont {Schaffer~P.}, \bibfnamefont {Kosasih~E.}},\ }\href@noop {} {\bibinfo {title} {Extracting supply chain knowledge graphs from natural language text using artificial intelligence}},\ \bibinfo {howpublished} {under review} (\bibinfo {year} {2023})\BibitemShut {NoStop}%
\bibitem [{\citenamefont {Reisch}\ \emph {et~al.}(2022)\citenamefont {Reisch}, \citenamefont {Heiler}, \citenamefont {Diem}, \citenamefont {Klimek},\ and\ \citenamefont {Thurner}}]{reisch2022inferring}%
  \BibitemOpen
  \bibfield  {author} {\bibinfo {author} {\bibfnamefont {T.}~\bibnamefont {Reisch}}, \bibinfo {author} {\bibfnamefont {G.}~\bibnamefont {Heiler}}, \bibinfo {author} {\bibfnamefont {C.}~\bibnamefont {Diem}}, \bibinfo {author} {\bibfnamefont {P.}~\bibnamefont {Klimek}},\ and\ \bibinfo {author} {\bibfnamefont {S.}~\bibnamefont {Thurner}},\ }\bibfield  {title} {\bibinfo {title} {{Monitoring supply networks from mobile phone data for estimating the systemic risk of an economy}},\ }\href {https://doi.org/10.1038/s41598-022-13104-5} {\bibfield  {journal} {\bibinfo  {journal} {Scientific Reports}\ }\textbf {\bibinfo {volume} {12}},\ \bibinfo {pages} {13347} (\bibinfo {year} {2022})},\ \bibinfo {note} {number: 1, Publisher: Nature Publishing Group}\BibitemShut {NoStop}%
\bibitem [{\citenamefont {Hooijmaaijers}\ and\ \citenamefont {Buiten}(2019)}]{Hooijmaaijers2019}%
  \BibitemOpen
  \bibfield  {author} {\bibinfo {author} {\bibfnamefont {S.}~\bibnamefont {Hooijmaaijers}}\ and\ \bibinfo {author} {\bibfnamefont {G.}~\bibnamefont {Buiten}},\ }\href {https://www.oecd.org/naec/new-economic-policymaking/Estimating_Interfirm_Trade_Network_Hooijmaaijers_Buiten.pdf} {\emph {\bibinfo {title} {A methodology for estimating the Dutch interfirm trade network, including a breakdown by commodity}}},\ \bibinfo {type} {Tech. Rep.}\ (\bibinfo  {institution} {Technical report, Statistics Netherlands},\ \bibinfo {year} {2019})\BibitemShut {NoStop}%
\bibitem [{\citenamefont {Hillman}\ \emph {et~al.}(2021)\citenamefont {Hillman}, \citenamefont {Barnes}, \citenamefont {Wharf},\ and\ \citenamefont {MacDonald}}]{hillman2021}%
  \BibitemOpen
  \bibfield  {author} {\bibinfo {author} {\bibfnamefont {R.}~\bibnamefont {Hillman}}, \bibinfo {author} {\bibfnamefont {S.}~\bibnamefont {Barnes}}, \bibinfo {author} {\bibfnamefont {G.}~\bibnamefont {Wharf}},\ and\ \bibinfo {author} {\bibfnamefont {D.}~\bibnamefont {MacDonald}},\ }\href {https://doi.org/https://doi.org/https://doi.org/10.1787/e9de0097-en} {\emph {\bibinfo {title} {A new firm-level model of corporate sector interactions and fragility: The Corporate Agent-Based (CAB) model}}},\ \bibinfo {type} {Tech. Rep.}\ \bibinfo {number} {1675}\ (\bibinfo  {institution} {OECD},\ \bibinfo {year} {2021})\BibitemShut {NoStop}%
\bibitem [{\citenamefont {Mungo}\ and\ \citenamefont {Moran}(2023)}]{mungo2023revealing}%
  \BibitemOpen
  \bibfield  {author} {\bibinfo {author} {\bibfnamefont {L.}~\bibnamefont {Mungo}}\ and\ \bibinfo {author} {\bibfnamefont {J.}~\bibnamefont {Moran}},\ }\href@noop {} {\bibinfo {title} {Revealing production networks from firm growth dynamics}} (\bibinfo {year} {2023}),\ \Eprint {https://arxiv.org/abs/2302.09906} {arXiv:2302.09906 [q-fin.ST]} \BibitemShut {NoStop}%
\bibitem [{\citenamefont {Hamilton}(2020)}]{hamilton2020}%
  \BibitemOpen
  \bibfield  {author} {\bibinfo {author} {\bibfnamefont {W.~L.}\ \bibnamefont {Hamilton}},\ }\href {https://doi.org/https://doi.org/10.1007/978-3-031-01588-5} {\emph {\bibinfo {title} {Graph Representation Learning}}},\ Synthesis Lectures on Artificial Intelligence and Machine Learning\ (\bibinfo  {publisher} {Springer Nature Switzerland},\ \bibinfo {year} {2020})\ p.\ \bibinfo {pages} {141}\BibitemShut {NoStop}%
\bibitem [{\citenamefont {Le}\ and\ \citenamefont {Mikolov}(2014)}]{doc2vec}%
  \BibitemOpen
  \bibfield  {author} {\bibinfo {author} {\bibfnamefont {Q.}~\bibnamefont {Le}}\ and\ \bibinfo {author} {\bibfnamefont {T.}~\bibnamefont {Mikolov}},\ }\bibfield  {title} {\bibinfo {title} {Distributed representations of sentences and documents},\ }in\ \href {https://proceedings.mlr.press/v32/le14.html} {\emph {\bibinfo {booktitle} {Proceedings of the 31st International Conference on Machine Learning}}},\ \bibinfo {series} {Proceedings of Machine Learning Research}, Vol.~\bibinfo {volume} {32},\ \bibinfo {editor} {edited by\ \bibinfo {editor} {\bibfnamefont {E.~P.}\ \bibnamefont {Xing}}\ and\ \bibinfo {editor} {\bibfnamefont {T.}~\bibnamefont {Jebara}}}\ (\bibinfo  {publisher} {PMLR},\ \bibinfo {address} {Bejing, China},\ \bibinfo {year} {2014})\ pp.\ \bibinfo {pages} {1188--1196}\BibitemShut {NoStop}%
\bibitem [{\citenamefont {Bernard}\ \emph {et~al.}(2022)\citenamefont {Bernard}, \citenamefont {Dhyne}, \citenamefont {Magerman}, \citenamefont {Manova},\ and\ \citenamefont {Moxnes}}]{bernard2019}%
  \BibitemOpen
  \bibfield  {author} {\bibinfo {author} {\bibfnamefont {A.~B.}\ \bibnamefont {Bernard}}, \bibinfo {author} {\bibfnamefont {E.}~\bibnamefont {Dhyne}}, \bibinfo {author} {\bibfnamefont {G.}~\bibnamefont {Magerman}}, \bibinfo {author} {\bibfnamefont {K.}~\bibnamefont {Manova}},\ and\ \bibinfo {author} {\bibfnamefont {A.}~\bibnamefont {Moxnes}},\ }\bibfield  {title} {\bibinfo {title} {The origins of firm heterogeneity: A production network approach},\ }\href {https://doi.org/10.1086/719759} {\bibfield  {journal} {\bibinfo  {journal} {Journal of Political Economy}\ }\textbf {\bibinfo {volume} {130}},\ \bibinfo {pages} {1765} (\bibinfo {year} {2022})},\ \Eprint {https://arxiv.org/abs/https://doi.org/10.1086/719759} {https://doi.org/10.1086/719759} \BibitemShut {NoStop}%
\bibitem [{\citenamefont {Rachkov}\ \emph {et~al.}(2021)\citenamefont {Rachkov}, \citenamefont {Pijpers},\ and\ \citenamefont {Garlaschelli}}]{rachkov2021}%
  \BibitemOpen
  \bibfield  {author} {\bibinfo {author} {\bibfnamefont {A.}~\bibnamefont {Rachkov}}, \bibinfo {author} {\bibfnamefont {F.}~\bibnamefont {Pijpers}},\ and\ \bibinfo {author} {\bibfnamefont {D.}~\bibnamefont {Garlaschelli}},\ }\href {https://www.cbs.nl/en-gb/background/2021/09/potential-biases-in-network-reconstruction-methods} {\emph {\bibinfo {title} {Potential Biases in Network Reconstruction Methods Not Maximizing Entropy}}},\ \bibinfo {type} {Tech. Rep.}\ (\bibinfo  {institution} {Discussion Papers, Central Bureau of Statistics, Netherlands},\ \bibinfo {year} {2021})\BibitemShut {NoStop}%
\bibitem [{\citenamefont {Watanabe}\ \emph {et~al.}(2013)\citenamefont {Watanabe}, \citenamefont {Takayasu},\ and\ \citenamefont {Takayasu}}]{watanabe2013relations}%
  \BibitemOpen
  \bibfield  {author} {\bibinfo {author} {\bibfnamefont {H.}~\bibnamefont {Watanabe}}, \bibinfo {author} {\bibfnamefont {H.}~\bibnamefont {Takayasu}},\ and\ \bibinfo {author} {\bibfnamefont {M.}~\bibnamefont {Takayasu}},\ }\bibfield  {title} {\bibinfo {title} {Relations between allometric scalings and fluctuations in complex systems: The case of japanese firms},\ }\href@noop {} {\bibfield  {journal} {\bibinfo  {journal} {Physica A: Statistical Mechanics and its Applications}\ }\textbf {\bibinfo {volume} {392}},\ \bibinfo {pages} {741} (\bibinfo {year} {2013})}\BibitemShut {NoStop}%
\bibitem [{\citenamefont {Parisi}\ \emph {et~al.}(2020)\citenamefont {Parisi}, \citenamefont {Squartini},\ and\ \citenamefont {Garlaschelli}}]{parisi2020}%
  \BibitemOpen
  \bibfield  {author} {\bibinfo {author} {\bibfnamefont {F.}~\bibnamefont {Parisi}}, \bibinfo {author} {\bibfnamefont {T.}~\bibnamefont {Squartini}},\ and\ \bibinfo {author} {\bibfnamefont {D.}~\bibnamefont {Garlaschelli}},\ }\bibfield  {title} {\bibinfo {title} {A faster horse on a safer trail: generalized inference for the efficient reconstruction of weighted networks},\ }\href {https://doi.org/10.1088/1367-2630/ab74a7} {\bibfield  {journal} {\bibinfo  {journal} {New Journal of Physics}\ }\textbf {\bibinfo {volume} {22}},\ \bibinfo {pages} {053053} (\bibinfo {year} {2020})}\BibitemShut {NoStop}%
\bibitem [{\citenamefont {Hazan}(2019)}]{hazan2019maximum}%
  \BibitemOpen
  \bibfield  {author} {\bibinfo {author} {\bibfnamefont {A.}~\bibnamefont {Hazan}},\ }\bibfield  {title} {\bibinfo {title} {A maximum entropy network reconstruction of macroeconomic models},\ }\href@noop {} {\bibfield  {journal} {\bibinfo  {journal} {Physica A: Statistical Mechanics and its Applications}\ }\textbf {\bibinfo {volume} {519}},\ \bibinfo {pages} {1} (\bibinfo {year} {2019})}\BibitemShut {NoStop}%
\bibitem [{\citenamefont {Garlaschelli}\ and\ \citenamefont {Loffredo}(2004)}]{garlaschelli_loffredo_2004}%
  \BibitemOpen
  \bibfield  {author} {\bibinfo {author} {\bibfnamefont {D.}~\bibnamefont {Garlaschelli}}\ and\ \bibinfo {author} {\bibfnamefont {M.~I.}\ \bibnamefont {Loffredo}},\ }\bibfield  {title} {\bibinfo {title} {Fitness-dependent topological properties of the world trade web},\ }\href {https://doi.org/10.1103/PhysRevLett.93.188701} {\bibfield  {journal} {\bibinfo  {journal} {Phys. Rev. Lett.}\ }\textbf {\bibinfo {volume} {93}},\ \bibinfo {pages} {188701} (\bibinfo {year} {2004})}\BibitemShut {NoStop}%
\bibitem [{\citenamefont {Dong}\ \emph {et~al.}(2019)\citenamefont {Dong}, \citenamefont {Thanou}, \citenamefont {Rabbat},\ and\ \citenamefont {Frossard}}]{dong2019}%
  \BibitemOpen
  \bibfield  {author} {\bibinfo {author} {\bibfnamefont {X.}~\bibnamefont {Dong}}, \bibinfo {author} {\bibfnamefont {D.}~\bibnamefont {Thanou}}, \bibinfo {author} {\bibfnamefont {M.}~\bibnamefont {Rabbat}},\ and\ \bibinfo {author} {\bibfnamefont {P.}~\bibnamefont {Frossard}},\ }\bibfield  {title} {\bibinfo {title} {Learning graphs from data: A signal representation perspective},\ }\href {https://doi.org/10.1109/MSP.2018.2887284} {\bibfield  {journal} {\bibinfo  {journal} {IEEE Signal Processing Magazine}\ }\textbf {\bibinfo {volume} {36}},\ \bibinfo {pages} {44} (\bibinfo {year} {2019})}\BibitemShut {NoStop}%
\bibitem [{\citenamefont {Peel}\ \emph {et~al.}(2022)\citenamefont {Peel}, \citenamefont {Peixoto},\ and\ \citenamefont {De~Domenico}}]{peixoto2022}%
  \BibitemOpen
  \bibfield  {author} {\bibinfo {author} {\bibfnamefont {L.}~\bibnamefont {Peel}}, \bibinfo {author} {\bibfnamefont {T.~P.}\ \bibnamefont {Peixoto}},\ and\ \bibinfo {author} {\bibfnamefont {M.}~\bibnamefont {De~Domenico}},\ }\bibfield  {title} {\bibinfo {title} {Statistical inference links data and theory in network science},\ }\href@noop {} {\bibfield  {journal} {\bibinfo  {journal} {Nature Communications}\ } (\bibinfo {year} {2022})}\BibitemShut {NoStop}%
\bibitem [{\citenamefont {Barrot}\ and\ \citenamefont {Sauvagnat}(2016)}]{barrot2016}%
  \BibitemOpen
  \bibfield  {author} {\bibinfo {author} {\bibfnamefont {J.-N.}\ \bibnamefont {Barrot}}\ and\ \bibinfo {author} {\bibfnamefont {J.}~\bibnamefont {Sauvagnat}},\ }\bibfield  {title} {\bibinfo {title} {{ Input Specificity and the Propagation of Idiosyncratic Shocks in Production Networks *}},\ }\href {https://doi.org/10.1093/qje/qjw018} {\bibfield  {journal} {\bibinfo  {journal} {The Quarterly Journal of Economics}\ }\textbf {\bibinfo {volume} {131}},\ \bibinfo {pages} {1543} (\bibinfo {year} {2016})},\ \Eprint {https://arxiv.org/abs/https://academic.oup.com/qje/article-pdf/131/3/1543/30636709/qjw018.pdf} {https://academic.oup.com/qje/article-pdf/131/3/1543/30636709/qjw018.pdf} \BibitemShut {NoStop}%
\bibitem [{\citenamefont {Carvalho}\ and\ \citenamefont {Tahbaz-Salehi}(2019)}]{Carvalho2019}%
  \BibitemOpen
  \bibfield  {author} {\bibinfo {author} {\bibfnamefont {V.~M.}\ \bibnamefont {Carvalho}}\ and\ \bibinfo {author} {\bibfnamefont {A.}~\bibnamefont {Tahbaz-Salehi}},\ }\bibfield  {title} {\bibinfo {title} {Production networks: A primer},\ }\href {https://doi.org/10.1146/annurev-economics-080218-030212} {\bibfield  {journal} {\bibinfo  {journal} {Annual Review of Economics}\ }\textbf {\bibinfo {volume} {11}},\ \bibinfo {pages} {635} (\bibinfo {year} {2019})}\BibitemShut {NoStop}%
\bibitem [{\citenamefont {Carvalho}\ \emph {et~al.}(2021)\citenamefont {Carvalho}, \citenamefont {Nirei}, \citenamefont {Saito},\ and\ \citenamefont {Tahbaz-Salehi}}]{carvalho2021}%
  \BibitemOpen
  \bibfield  {author} {\bibinfo {author} {\bibfnamefont {V.~M.}\ \bibnamefont {Carvalho}}, \bibinfo {author} {\bibfnamefont {M.}~\bibnamefont {Nirei}}, \bibinfo {author} {\bibfnamefont {Y.~U.}\ \bibnamefont {Saito}},\ and\ \bibinfo {author} {\bibfnamefont {A.}~\bibnamefont {Tahbaz-Salehi}},\ }\bibfield  {title} {\bibinfo {title} {{Supply Chain Disruptions: Evidence from the Great East Japan Earthquake}},\ }\href {https://doi.org/10.1093/qje/qjaa044} {\bibfield  {journal} {\bibinfo  {journal} {The Quarterly Journal of Economics}\ }\textbf {\bibinfo {volume} {136}},\ \bibinfo {pages} {1255} (\bibinfo {year} {2021})},\ \Eprint {https://arxiv.org/abs/https://academic.oup.com/qje/article-pdf/136/2/1255/36725306/qjaa044.pdf} {https://academic.oup.com/qje/article-pdf/136/2/1255/36725306/qjaa044.pdf} \BibitemShut {NoStop}%
\bibitem [{\citenamefont {Kumar}\ \emph {et~al.}(2019)\citenamefont {Kumar}, \citenamefont {Ying}, \citenamefont {de~Miranda~Cardoso},\ and\ \citenamefont {Palomar}}]{Kumar_et_al_2019}%
  \BibitemOpen
  \bibfield  {author} {\bibinfo {author} {\bibfnamefont {S.}~\bibnamefont {Kumar}}, \bibinfo {author} {\bibfnamefont {J.}~\bibnamefont {Ying}}, \bibinfo {author} {\bibfnamefont {J.~V.}\ \bibnamefont {de~Miranda~Cardoso}},\ and\ \bibinfo {author} {\bibfnamefont {D.}~\bibnamefont {Palomar}},\ }\bibfield  {title} {\bibinfo {title} {Structured graph learning via laplacian spectral constraints},\ }in\ \href {https://proceedings.neurips.cc/paper/2019/file/90cc440b1b8caa520c562ac4e4bbcb51-Paper.pdf} {\emph {\bibinfo {booktitle} {Advances in Neural Information Processing Systems}}},\ Vol.~\bibinfo {volume} {32},\ \bibinfo {editor} {edited by\ \bibinfo {editor} {\bibfnamefont {H.}~\bibnamefont {Wallach}}, \bibinfo {editor} {\bibfnamefont {H.}~\bibnamefont {Larochelle}}, \bibinfo {editor} {\bibfnamefont {A.}~\bibnamefont {Beygelzimer}}, \bibinfo {editor} {\bibfnamefont {F.}~\bibnamefont {Alch\'{e}-Buc}}, \bibinfo {editor} {\bibfnamefont {E.}~\bibnamefont {Fox}},\ and\ \bibinfo {editor} {\bibfnamefont {R.}~\bibnamefont
  {Garnett}}}\ (\bibinfo  {publisher} {Curran Associates, Inc.},\ \bibinfo {year} {2019})\BibitemShut {NoStop}%
\bibitem [{\citenamefont {Anand}\ \emph {et~al.}(2018)\citenamefont {Anand}, \citenamefont {{van Lelyveld}}, \citenamefont {Ádám Banai}, \citenamefont {Friedrich}, \citenamefont {Garratt}, \citenamefont {Hałaj}, \citenamefont {Fique}, \citenamefont {Hansen}, \citenamefont {Jaramillo}, \citenamefont {Lee}, \citenamefont {Molina-Borboa}, \citenamefont {Nobili}, \citenamefont {Rajan}, \citenamefont {Salakhova}, \citenamefont {Silva}, \citenamefont {Silvestri},\ and\ \citenamefont {{de Souza}}}]{anand2018}%
  \BibitemOpen
  \bibfield  {author} {\bibinfo {author} {\bibfnamefont {K.}~\bibnamefont {Anand}}, \bibinfo {author} {\bibfnamefont {I.}~\bibnamefont {{van Lelyveld}}}, \bibinfo {author} {\bibnamefont {Ádám Banai}}, \bibinfo {author} {\bibfnamefont {S.}~\bibnamefont {Friedrich}}, \bibinfo {author} {\bibfnamefont {R.}~\bibnamefont {Garratt}}, \bibinfo {author} {\bibfnamefont {G.}~\bibnamefont {Hałaj}}, \bibinfo {author} {\bibfnamefont {J.}~\bibnamefont {Fique}}, \bibinfo {author} {\bibfnamefont {I.}~\bibnamefont {Hansen}}, \bibinfo {author} {\bibfnamefont {S.~M.}\ \bibnamefont {Jaramillo}}, \bibinfo {author} {\bibfnamefont {H.}~\bibnamefont {Lee}}, \bibinfo {author} {\bibfnamefont {J.~L.}\ \bibnamefont {Molina-Borboa}}, \bibinfo {author} {\bibfnamefont {S.}~\bibnamefont {Nobili}}, \bibinfo {author} {\bibfnamefont {S.}~\bibnamefont {Rajan}}, \bibinfo {author} {\bibfnamefont {D.}~\bibnamefont {Salakhova}}, \bibinfo {author} {\bibfnamefont {T.~C.}\ \bibnamefont {Silva}}, \bibinfo {author} {\bibfnamefont {L.}~\bibnamefont
  {Silvestri}},\ and\ \bibinfo {author} {\bibfnamefont {S.~R.~S.}\ \bibnamefont {{de Souza}}},\ }\bibfield  {title} {\bibinfo {title} {The missing links: A global study on uncovering financial network structures from partial data},\ }\href {https://doi.org/https://doi.org/10.1016/j.jfs.2017.05.012} {\bibfield  {journal} {\bibinfo  {journal} {Journal of Financial Stability}\ }\textbf {\bibinfo {volume} {35}},\ \bibinfo {pages} {107} (\bibinfo {year} {2018})},\ \bibinfo {note} {network models, stress testing and other tools for financial stability monitoring and macroprudential policy design and implementation}\BibitemShut {NoStop}%
\bibitem [{\citenamefont {Cimini}\ \emph {et~al.}(2021)\citenamefont {Cimini}, \citenamefont {Mastrandrea},\ and\ \citenamefont {Squartini}}]{cimini_mastrandrea_squartini_2021}%
  \BibitemOpen
  \bibfield  {author} {\bibinfo {author} {\bibfnamefont {G.}~\bibnamefont {Cimini}}, \bibinfo {author} {\bibfnamefont {R.}~\bibnamefont {Mastrandrea}},\ and\ \bibinfo {author} {\bibfnamefont {T.}~\bibnamefont {Squartini}},\ }\href {https://doi.org/10.1017/9781108771030} {\emph {\bibinfo {title} {Reconstructing Networks}}},\ Elements in the Structure and Dynamics of Complex Networks\ (\bibinfo  {publisher} {Cambridge University Press},\ \bibinfo {year} {2021})\BibitemShut {NoStop}%
\bibitem [{\citenamefont {Golan}\ \emph {et~al.}(1994)\citenamefont {Golan}, \citenamefont {Judge},\ and\ \citenamefont {Robinson}}]{golan1994}%
  \BibitemOpen
  \bibfield  {author} {\bibinfo {author} {\bibfnamefont {A.}~\bibnamefont {Golan}}, \bibinfo {author} {\bibfnamefont {G.}~\bibnamefont {Judge}},\ and\ \bibinfo {author} {\bibfnamefont {S.}~\bibnamefont {Robinson}},\ }\bibfield  {title} {\bibinfo {title} {Recovering information from incomplete or partial multisectoral economic data},\ }\href {http://www.jstor.org/stable/2109978} {\bibfield  {journal} {\bibinfo  {journal} {The Review of Economics and Statistics}\ }\textbf {\bibinfo {volume} {76}},\ \bibinfo {pages} {541} (\bibinfo {year} {1994})}\BibitemShut {NoStop}%
\bibitem [{\citenamefont {Inoue}\ and\ \citenamefont {Todo}(2019)}]{InoueTodo2019}%
  \BibitemOpen
  \bibfield  {author} {\bibinfo {author} {\bibfnamefont {H.}~\bibnamefont {Inoue}}\ and\ \bibinfo {author} {\bibfnamefont {Y.}~\bibnamefont {Todo}},\ }\bibfield  {title} {\bibinfo {title} {Firm-level propagation of shocks through supply-chain networks},\ }\href {https://doi.org/10.1038/s41893-019-0351-x} {\bibfield  {journal} {\bibinfo  {journal} {Nature Sustainability}\ }\textbf {\bibinfo {volume} {2}},\ \bibinfo {pages} {841} (\bibinfo {year} {2019})}\BibitemShut {NoStop}%
\bibitem [{\citenamefont {Welburn}\ \emph {et~al.}(2020)\citenamefont {Welburn}, \citenamefont {Strong}, \citenamefont {Nekoul}, \citenamefont {Grana}, \citenamefont {Marcinek}, \citenamefont {Osoba}, \citenamefont {Koirala},\ and\ \citenamefont {Setodji}}]{welburn2020}%
  \BibitemOpen
  \bibfield  {author} {\bibinfo {author} {\bibfnamefont {J.~W.}\ \bibnamefont {Welburn}}, \bibinfo {author} {\bibfnamefont {A.}~\bibnamefont {Strong}}, \bibinfo {author} {\bibfnamefont {F.~E.}\ \bibnamefont {Nekoul}}, \bibinfo {author} {\bibfnamefont {J.}~\bibnamefont {Grana}}, \bibinfo {author} {\bibfnamefont {K.}~\bibnamefont {Marcinek}}, \bibinfo {author} {\bibfnamefont {O.~A.}\ \bibnamefont {Osoba}}, \bibinfo {author} {\bibfnamefont {N.}~\bibnamefont {Koirala}},\ and\ \bibinfo {author} {\bibfnamefont {C.~M.}\ \bibnamefont {Setodji}},\ }\href {https://doi.org/10.7249/RR4185} {\emph {\bibinfo {title} {Systemic Risk in the Broad Economy: Interfirm Networks and Shocks in the U.S. Economy}}}\ (\bibinfo  {publisher} {RAND Corporation},\ \bibinfo {address} {Santa Monica, CA},\ \bibinfo {year} {2020})\BibitemShut {NoStop}%
\bibitem [{\citenamefont {Bacilieri}\ and\ \citenamefont {Astudillo-Estevez}(2023)}]{bacilieri2023reconstructing}%
  \BibitemOpen
  \bibfield  {author} {\bibinfo {author} {\bibfnamefont {A.}~\bibnamefont {Bacilieri}}\ and\ \bibinfo {author} {\bibfnamefont {P.}~\bibnamefont {Astudillo-Estevez}},\ }\href@noop {} {\bibinfo {title} {Reconstructing firm-level input-output networks from partial information}} (\bibinfo {year} {2023}),\ \Eprint {https://arxiv.org/abs/2304.00081} {arXiv:2304.00081 [econ.GN]} \BibitemShut {NoStop}%
\bibitem [{\citenamefont {Jaynes}(1957)}]{jaynes1957}%
  \BibitemOpen
  \bibfield  {author} {\bibinfo {author} {\bibfnamefont {E.~T.}\ \bibnamefont {Jaynes}},\ }\bibfield  {title} {\bibinfo {title} {Information theory and statistical mechanics},\ }\href {https://doi.org/10.1103/PhysRev.106.620} {\bibfield  {journal} {\bibinfo  {journal} {Phys. Rev.}\ }\textbf {\bibinfo {volume} {106}},\ \bibinfo {pages} {620} (\bibinfo {year} {1957})}\BibitemShut {NoStop}%
\bibitem [{\citenamefont {Bartolucci}\ \emph {et~al.}(2023)\citenamefont {Bartolucci}, \citenamefont {Caccioli}, \citenamefont {Caravelli},\ and\ \citenamefont {Vivo}}]{bartolucci2023ranking}%
  \BibitemOpen
  \bibfield  {author} {\bibinfo {author} {\bibfnamefont {S.}~\bibnamefont {Bartolucci}}, \bibinfo {author} {\bibfnamefont {F.}~\bibnamefont {Caccioli}}, \bibinfo {author} {\bibfnamefont {F.}~\bibnamefont {Caravelli}},\ and\ \bibinfo {author} {\bibfnamefont {P.}~\bibnamefont {Vivo}},\ }\bibfield  {title} {\bibinfo {title} {Ranking influential nodes in networks from aggregate local information},\ }\href@noop {} {\bibfield  {journal} {\bibinfo  {journal} {Physical Review Research}\ }\textbf {\bibinfo {volume} {5}},\ \bibinfo {pages} {033123} (\bibinfo {year} {2023})}\BibitemShut {NoStop}%
\bibitem [{\citenamefont {Bartolucci}\ \emph {et~al.}(2020)\citenamefont {Bartolucci}, \citenamefont {Caccioli}, \citenamefont {Caravelli},\ and\ \citenamefont {Vivo}}]{bartolucci2020}%
  \BibitemOpen
  \bibfield  {author} {\bibinfo {author} {\bibfnamefont {S.}~\bibnamefont {Bartolucci}}, \bibinfo {author} {\bibfnamefont {F.}~\bibnamefont {Caccioli}}, \bibinfo {author} {\bibfnamefont {F.}~\bibnamefont {Caravelli}},\ and\ \bibinfo {author} {\bibfnamefont {P.}~\bibnamefont {Vivo}},\ }\href {https://doi.org/https://doi.org/10.48550/arXiv.2009.06350} {\bibinfo {title} {Inversion-free leontief inverse: statistical regularities in input-output analysis from partial information}} (\bibinfo {year} {2020})\BibitemShut {NoStop}%
\bibitem [{\citenamefont {Acemoglu}\ \emph {et~al.}(2012)\citenamefont {Acemoglu}, \citenamefont {Carvalho}, \citenamefont {Ozdaglar},\ and\ \citenamefont {Tahbaz-Salehi}}]{acemoglu2012network}%
  \BibitemOpen
  \bibfield  {author} {\bibinfo {author} {\bibfnamefont {D.}~\bibnamefont {Acemoglu}}, \bibinfo {author} {\bibfnamefont {V.}~\bibnamefont {Carvalho}}, \bibinfo {author} {\bibfnamefont {A.}~\bibnamefont {Ozdaglar}},\ and\ \bibinfo {author} {\bibfnamefont {A.}~\bibnamefont {Tahbaz-Salehi}},\ }\bibfield  {title} {\bibinfo {title} {The network origins of aggregate fluctuations},\ }\href {https://doi.org/10.3982/ecta9623} {\bibfield  {journal} {\bibinfo  {journal} {Econometrica}\ }\textbf {\bibinfo {volume} {80}},\ \bibinfo {pages} {1977} (\bibinfo {year} {2012})}\BibitemShut {NoStop}%
\bibitem [{\citenamefont {Occhini}\ \emph {et~al.}(2023)\citenamefont {Occhini}, \citenamefont {Tranos},\ and\ \citenamefont {Wolf}}]{occhini2023measuring}%
  \BibitemOpen
  \bibfield  {author} {\bibinfo {author} {\bibfnamefont {G.}~\bibnamefont {Occhini}}, \bibinfo {author} {\bibfnamefont {E.}~\bibnamefont {Tranos}},\ and\ \bibinfo {author} {\bibfnamefont {L.}~\bibnamefont {Wolf}},\ }\href@noop {} {\bibinfo {title} {Measuring a country’s digital industrial structure: commercial websites and weakly supervised classification to the rescue}},\ \bibinfo {howpublished} {SocArXiv} (\bibinfo {year} {2023})\BibitemShut {NoStop}%
\bibitem [{\citenamefont {Atalay}\ \emph {et~al.}(2011)\citenamefont {Atalay}, \citenamefont {Hortacsu}, \citenamefont {Roberts},\ and\ \citenamefont {Syverson}}]{AtalayEtAl2011}%
  \BibitemOpen
  \bibfield  {author} {\bibinfo {author} {\bibfnamefont {E.}~\bibnamefont {Atalay}}, \bibinfo {author} {\bibfnamefont {A.}~\bibnamefont {Hortacsu}}, \bibinfo {author} {\bibfnamefont {J.}~\bibnamefont {Roberts}},\ and\ \bibinfo {author} {\bibfnamefont {C.}~\bibnamefont {Syverson}},\ }\bibfield  {title} {\bibinfo {title} {Network structure of production},\ }\href {https://doi.org/https://doi.org/10.1073/pnas.1015564108} {\bibfield  {journal} {\bibinfo  {journal} {Proceedings of the National Academy of Sciences}\ }\textbf {\bibinfo {volume} {108}},\ \bibinfo {pages} {5199} (\bibinfo {year} {2011})}\BibitemShut {NoStop}%
\bibitem [{\citenamefont {Russakovsky}\ \emph {et~al.}(2015)\citenamefont {Russakovsky}, \citenamefont {Deng}, \citenamefont {Su}, \citenamefont {Krause}, \citenamefont {Satheesh}, \citenamefont {Ma}, \citenamefont {Huang}, \citenamefont {Karpathy}, \citenamefont {Khosla}, \citenamefont {Bernstein}, \citenamefont {Berg},\ and\ \citenamefont {Fei-Fei}}]{imagenet2015}%
  \BibitemOpen
  \bibfield  {author} {\bibinfo {author} {\bibfnamefont {O.}~\bibnamefont {Russakovsky}}, \bibinfo {author} {\bibfnamefont {J.}~\bibnamefont {Deng}}, \bibinfo {author} {\bibfnamefont {H.}~\bibnamefont {Su}}, \bibinfo {author} {\bibfnamefont {J.}~\bibnamefont {Krause}}, \bibinfo {author} {\bibfnamefont {S.}~\bibnamefont {Satheesh}}, \bibinfo {author} {\bibfnamefont {S.}~\bibnamefont {Ma}}, \bibinfo {author} {\bibfnamefont {Z.}~\bibnamefont {Huang}}, \bibinfo {author} {\bibfnamefont {A.}~\bibnamefont {Karpathy}}, \bibinfo {author} {\bibfnamefont {A.}~\bibnamefont {Khosla}}, \bibinfo {author} {\bibfnamefont {M.}~\bibnamefont {Bernstein}}, \bibinfo {author} {\bibfnamefont {A.~C.}\ \bibnamefont {Berg}},\ and\ \bibinfo {author} {\bibfnamefont {L.}~\bibnamefont {Fei-Fei}},\ }\bibfield  {title} {\bibinfo {title} {{ImageNet} {Large} {Scale} {Visual} {Recognition} {Challenge}},\ }\href {https://doi.org/10.1007/s11263-015-0816-y} {\bibfield  {journal} {\bibinfo  {journal} {International Journal of Computer Vision}\
  }\textbf {\bibinfo {volume} {115}},\ \bibinfo {pages} {211} (\bibinfo {year} {2015})}\BibitemShut {NoStop}%
\bibitem [{\citenamefont {Deng}\ \emph {et~al.}(2009)\citenamefont {Deng}, \citenamefont {Dong}, \citenamefont {Socher}, \citenamefont {Li}, \citenamefont {Li},\ and\ \citenamefont {Fei-Fei}}]{imagenet2009}%
  \BibitemOpen
  \bibfield  {author} {\bibinfo {author} {\bibfnamefont {J.}~\bibnamefont {Deng}}, \bibinfo {author} {\bibfnamefont {W.}~\bibnamefont {Dong}}, \bibinfo {author} {\bibfnamefont {R.}~\bibnamefont {Socher}}, \bibinfo {author} {\bibfnamefont {L.-J.}\ \bibnamefont {Li}}, \bibinfo {author} {\bibfnamefont {K.}~\bibnamefont {Li}},\ and\ \bibinfo {author} {\bibfnamefont {L.}~\bibnamefont {Fei-Fei}},\ }\bibfield  {title} {\bibinfo {title} {Imagenet: A large-scale hierarchical image database},\ }in\ \href {https://doi.org/10.1109/CVPR.2009.5206848} {\emph {\bibinfo {booktitle} {2009 IEEE Conference on Computer Vision and Pattern Recognition}}}\ (\bibinfo {year} {2009})\ pp.\ \bibinfo {pages} {248--255}\BibitemShut {NoStop}%
\bibitem [{\citenamefont {Jordon}\ \emph {et~al.}(2022)\citenamefont {Jordon}, \citenamefont {Szpruch}, \citenamefont {Houssiau}, \citenamefont {Bottarelli}, \citenamefont {Cherubin}, \citenamefont {Maple}, \citenamefont {Cohen},\ and\ \citenamefont {Weller}}]{synthetic2022}%
  \BibitemOpen
  \bibfield  {author} {\bibinfo {author} {\bibfnamefont {J.}~\bibnamefont {Jordon}}, \bibinfo {author} {\bibfnamefont {L.}~\bibnamefont {Szpruch}}, \bibinfo {author} {\bibfnamefont {F.}~\bibnamefont {Houssiau}}, \bibinfo {author} {\bibfnamefont {M.}~\bibnamefont {Bottarelli}}, \bibinfo {author} {\bibfnamefont {G.}~\bibnamefont {Cherubin}}, \bibinfo {author} {\bibfnamefont {C.}~\bibnamefont {Maple}}, \bibinfo {author} {\bibfnamefont {S.~N.}\ \bibnamefont {Cohen}},\ and\ \bibinfo {author} {\bibfnamefont {A.}~\bibnamefont {Weller}},\ }\href@noop {} {\bibinfo {title} {Synthetic data -- what, why and how?}} (\bibinfo {year} {2022}),\ \Eprint {https://arxiv.org/abs/2205.03257} {arXiv:2205.03257 [cs.LG]} \BibitemShut {NoStop}%
\bibitem [{\citenamefont {Ivanov}\ and\ \citenamefont {Dolgui}(2022)}]{ivanov2022shortage}%
  \BibitemOpen
  \bibfield  {author} {\bibinfo {author} {\bibfnamefont {D.}~\bibnamefont {Ivanov}}\ and\ \bibinfo {author} {\bibfnamefont {A.}~\bibnamefont {Dolgui}},\ }\bibfield  {title} {\bibinfo {title} {The shortage economy and its implications for supply chain and operations management},\ }\href@noop {} {\bibfield  {journal} {\bibinfo  {journal} {International Journal of Production Research}\ }\textbf {\bibinfo {volume} {60}},\ \bibinfo {pages} {7141} (\bibinfo {year} {2022})}\BibitemShut {NoStop}%
\bibitem [{\citenamefont {Brintrup}\ \emph {et~al.}(2022)\citenamefont {Brintrup}, \citenamefont {Kosasih}, \citenamefont {MacCarthy},\ and\ \citenamefont {Demirel}}]{brintrup2022digital}%
  \BibitemOpen
  \bibfield  {author} {\bibinfo {author} {\bibfnamefont {A.}~\bibnamefont {Brintrup}}, \bibinfo {author} {\bibfnamefont {E.~E.}\ \bibnamefont {Kosasih}}, \bibinfo {author} {\bibfnamefont {B.~L.}\ \bibnamefont {MacCarthy}},\ and\ \bibinfo {author} {\bibfnamefont {G.}~\bibnamefont {Demirel}},\ }\bibfield  {title} {\bibinfo {title} {Digital supply chain surveillance: concepts, challenges, and frameworks},\ }in\ \href@noop {} {\emph {\bibinfo {booktitle} {The Digital Supply Chain}}}\ (\bibinfo  {publisher} {Elsevier},\ \bibinfo {year} {2022})\ pp.\ \bibinfo {pages} {379--396}\BibitemShut {NoStop}%
\end{thebibliography}%


\end{document}